\documentclass[%
 reprint,
superscriptaddress,
 amsmath,amssymb,
 aps,nofootinbib
]{revtex4-2}

\usepackage{graphicx}
\usepackage{dcolumn}
\usepackage{bm}

\usepackage{physics}
\usepackage{slashed}
\usepackage{cleveref}
\usepackage{mathrsfs} 
\usepackage{booktabs}
\usepackage{multirow}
\usepackage{makecell}
\usepackage[caption=false]{subfig} 
\usepackage{longtable}
\usepackage{upgreek}
\usepackage{xcolor}
\usepackage{float}
\usepackage{tikz}
\usepackage{appendix}
\usepackage{ulem}

\newcolumntype{s}{>{\centering\arraybackslash}X}

\newcommand{\aerror}[2]{\substack{\scriptscriptstyle +#1\\ \scriptscriptstyle-#2}}

\begin{document}

\title{
A Three-Coupled-Channel Analysis of $Z_c(3900)$ Involving $D\bar{D}^*$, $\pi J/\psi$, and $\rho \eta_c $ 
}

\author{Kang Yu}
\email{yukang21@mails.ucas.ac.cn}
\affiliation{%
School of Physical Sciences, University of Chinese Academy of Sciences, Beijing 100049, China}
\author{Guang-Juan Wang}
\email{wgj@post.kek.jp}
\affiliation{
KEK Theory Center, Institute of Particle and Nuclear Studies (IPNS), High Energy Accelerator Research Organization (KEK), Tsukuba 305-0801, Japan
}
\author{Jia-Jun Wu}
\email{corresponding author, wujiajun@ucas.ac.cn}
\affiliation{School of Physical Sciences, University of Chinese Academy of Sciences, Beijing 100049, China}
\affiliation{Southern Center for Nuclear-Science Theory (SCNT), Institute of Modern Physics, Chinese Academy of Sciences, Huizhou 516000, China}\vspace{0.5cm}
\author{Zhi Yang}
\email{zhiyang@uestc.edu.cn}
\affiliation{School of Physics, University of Electronic Science and Technology of China, Chengdu 610054, China
}

\begin{abstract}
In this work, we conduct a three-coupled-channel analysis of the $Z_c(3900)$ structure, focusing on the $D\bar{D}^*$, $J/\psi \pi$, and $\rho \eta_c$ channels, based on the one-boson exchange model. Drawing from previous study on the exotic state $T_{cc}$, we only utilize one more parameter to construct the interactions between the channels. Our model successfully reproduces the experimental line shapes of the invariant mass distribution at $\sqrt{s} = 4.23$ and $4.26$ GeV for the three channels. Additionally, the finite-volume energy levels in our model show agreement with current LQCD conclusion.
Detailed analysis suggests that the $Z_c(3900)$ peaks in the $\pi J/\psi$ and $\rho \eta_c$ distributions primarily arise from the triangle loop involving the $D_1 \bar{D} D^*$ intermediate system. In the $D\bar{D}^*$ distribution, the threshold peak is generated by the cascade decay mechanism enhanced by a triangle diagram. Moreover, we find a virtual pole located far from the threshold, indicating that $Z_c(3900)$ peaks are not associated with physical pole. We conclude that the $Z_c(3900)$ peaks are predominantly caused by the threshold cusp.


\end{abstract}

\maketitle

\section{Introduction}\label{sec:intro}

The discovery of the $X(3872)$ near the $D\bar{D}^*$ threshold in 2003 marked a new era in hadron physics~\cite{BelleX3872}. Since then, dozens of exotic hadrons, collectively referred to as the XYZ structures, have been discovered in the heavy quarkonium region. These states significantly challenge conventional quark models, as their masses or quantum numbers cannot be adequately described by traditional two-quark ($q\bar{q}$) or three-quark ($qqq$) configurations.
%
The charged state $Z_c(3900)^\pm$ was almost simultaneously observed by BESIII~\cite{BES3pj426} and Belle~\cite{BelleZc} in the $\pi^\pm J/\psi$ invariant mass spectrum  from the process  $e^+ e^- \to \pi^+\pi^- J/\psi$ at $\sqrt{s} = 4.26$ GeV. CLEO-c further confirmed the existence of the charged structure in the same channel at a slightly lower $e^+e^-$ collision energy $\sqrt{s}=4.17$ GeV.  They also reported observation of the  neutral partner, $Z_c(3900)^0$, in the process $e^+ e^- \to \pi^0\pi^0 J/\psi$~\cite{CLEOc}.

Subsequently, BESIII studied the open-charm mode $e^+e^-\to\pi^\pm (D\bar{D}^*)^\mp$ at both $\sqrt{s}=4.23$ and $4.26$ GeV, initially using a single-tag technique~\cite{BES3DD426} and later refining the analysis with a double-tag technique~\cite{BES3DD423}. In both studies, they observed a near-threshold peak structure at approximately $3.85$ GeV in the $(D\bar{D}^*)^\pm$ invariant mass distribution, identified as $Z_c(3885)^\pm$. The Breit-Wigner mass and width for $Z_c(3900)$ and $Z_c(3885)$, as measured by BESIII, are $(m,\Gamma) = (3894.5 \pm 6.6 \pm 4.5,63\pm24\pm26)$ MeV and $(m,\Gamma) = (3882.2\pm 1.1 \pm 1.5, 26.5 \pm 1.7 \pm 2.1)$ MeV, respectively.
Due to their similar widths, $Z_c(3885)^\pm$ and $Z_c(3900)^\pm$ are generally considered to be the same entity, despite the slight difference in their masses.\footnote{For simplicity, we will refer to them collectively as $Z_c$ unless a distinction is necessary. 
}
A partial-wave analysis suggests that the spin-parity quantum number $J^P$ is likely $1^+$~\cite{BES3pj423}.
Recently, a clear signal with a statistical significance exceeding $4\sigma$ was observed in the process $e^+e^- \to \pi^+\pi^0\pi^- \eta_c$ at $\sqrt{s}=4.23$ GeV. This was achieved by constraining  the invariant masses of the $\pi^\pm\pi^0$ and $\eta_c$ candidates to fall within the $\rho^\pm$-signal and $\eta_c$-signal regions, respectively~\cite{Yuan:2018inv}.
So far, the $Z_c(3900)$ has only been observed in $e^+e^-$ production.  No clear evidence of  $Z_c(3900)$ has been found in $B$-decay processes, such as $\bar{B}^0 \to (J/\psi \pi^+)K^-$ in Belle~\cite{Belle:2014nuw}, $B^0 \to (J/\psi \pi^+)\pi^-$ and $B^0 \to (\psi(nS)\pi^+)K^-$ in LHCb~\cite{ LHCb:2014vbo,Johnson:2024omq}.

Since the discovery of $Z_c$, extensive theoretical research has been conducted to elucidate its nature. The proposed explanations for the $Z_c$ structure include compact tetraquark states~\cite{Braaten:2013boa, Dias:2013xfa, Qiao:2013raa, Wang:2020iqt, Maiani:2021tri}, $D\bar{D}^*$ resonances or virtual molecular states~\cite{Wang:2013cya, Aceti:2014uea, Albaladejo:2015lob, Gong:2016hlt, He:2017lhy, Ortega:2018cnm, Yang:2020nrt, Yan:2021tcp, Du:2022jjv, Chen:2023def, Lin:2024qcq}, cusp effects~\cite{Swanson:2014tra}, and non-resonant mechanisms~\cite{Wang:2020axi, vonDetten:2024eie}. Despite this extensive research, the exact nature of $Z_c$ is still uncertain.
%
In experiments, $Z_c$ is suggested to be a typical resonance based on Breit-Wigner parameterizations. However, some theoretical studies suggest that the observed peak might not necessarily indicate a true resonance. The peak could instead be a virtual state or even a cusp effect, rather than a resonance, depending on the parameterization strategy used.

To reach a definitive conclusion, it is essential to impose further constraints on the model and parameters, either through experimental data or underlying symmetries. Therefore, it is essential for a comprehensive model to be capable of explaining as much experimental data as possible.
However, capturing the dynamics of the three coupled channels, $\pi J/\psi$, $D\bar{D}^*$, and $\rho\eta_c$, which predominantly govern the nature of $Z_c(3900)$, remains a challenge.  There is a pressing need for a model that can consistently describe the $Z_c(3900)$ peaks from different final channels.
%

The nature of the $Z_c$ structure has also been investigated by several lattice QCD(LQCD) collaborations. The first LQCD study of $Z_c$ was conducted in Ref.~\cite{Prelovsek:2013xba} using $\pi J/\psi$ and $D\bar{D}^*$ meson-meson interpolators at $m_\pi \approx 266$ MeV.  A subsequent study~\cite{Prelovsek:2014swa} included additional meson-meson and diquark-antidiquark interpolators were included in the simulation.
In Ref.~\cite{Cheung:2016bym}, the Hadron Spectrum Collaboration (HSC) found that compact tetraquark operators had minimal impact on finite volume spectra at $m_\pi\approx 391$ MeV. Similarly, the Chinese LQCD(CLQCD) collaboration conducted simulations using meson-meson interpolators at three different $m_\pi$ values~\cite{Chen:2014afa,Liu:2019gmh}. However, none of these studies identified a clear candidate for the $Z_c$ state among the energy levels extracted from their lattice configurations.
In Refs.~\cite{HALQCD:2016ofq, Ikeda:2017mee}, the HALQCD collaboration extracted the $\pi J/\psi - D\bar{D}^* - \rho\eta_c$ coupled-channel effective potential using their formalism and successfully reproduced the experimental line shape. They concluded that the $Z_c$ state is better understood as a threshold cusp rather than a conventional resonance~\cite{HALQCD:2016ofq, Ikeda:2017mee}. 
Besides, it was emphasized that the off-diagonal channel-channel interactions play an important role on $Z_c$ therefore the couple-channel analysis is necessary.


As previously discussed, a significant limitation of prior studies is the absence of a model that integrates the $D\bar{D}^*(+\bar{D}D^*)$ \footnote{Hereafter simplified as  $D\bar{D}^*$.}, $J/\psi \pi$, and $\eta_c \rho$ three coupled channels to construct a comprehensive picture of $Z_c$.
In this work, we employ the Hamiltonian Effective Field Theory (HEFT) to investigate the $Z_c$ state from both phenomenological and lattice perspectives, focusing on the $D\bar{D}^*-\pi J/\psi-\rho \eta_c$ three coupled channels, where $Z_c$ signals have been observed. The One-Boson-Exchange (OBE) interaction between $D\bar{D}^*$ channels, which respects heavy quark spin symmetry, is utilized here.
Building on previous research on the double-charmed exotic state $T_{cc}^+$ ~\cite{Wang:2023ovj}, the interaction between $D\bar{D}^*$ and $D\bar{D}^*$ is determined, leaving only one free coupling parameter for the vertex $J/\psi-D\bar{D}^*$ and $\eta_c-D\bar{D}^*$ to construct the full channel-channel interaction. This parameter is calibrated by simultaneously fitting experimental data across the three channels. 
The vertices like $Y\to\pi D\bar{D}^*,\,Y\to D_1\bar{D}$ are also more constrained in the couple-channel analysis compared to the single channel one.

%

By constraining our model and parameter in this way, we are able to provide a potential explanation for $Z_c$ which is also consistent with previous analyses of $T_{cc}$. 
With all parameters determined, we then calculate the finite volume energy levels and compare them with previous LQCD results. It will be shown later that our model provides an interpretation of $Z_c$ that is consistent with both experimental data and lattice results.
%


This paper is organized as follows. In Sec. \ref{sec:formalism}, we outline the formalism used for the fitting process. In Sec. \ref{sec:fitting} we present the fitting results, along with a discussion of the observed peak structures. In Sec. \ref{sec:pole}, we analytically continue the scattering matrix and search for its poles in the complex plane. In Sec.~\ref{sec:finitevol}, we provide a brief introduction to the HEFT and applies it to calculate the finite volume energy levels, which are then compared to LQCD results. Finally, in Sec.~\ref{sec:summary}, we make a summary and discuss future prospects.

\section{Formalism}\label{sec:formalism}


\begin{figure*}
    \centering
    \includegraphics[width=0.95\linewidth]{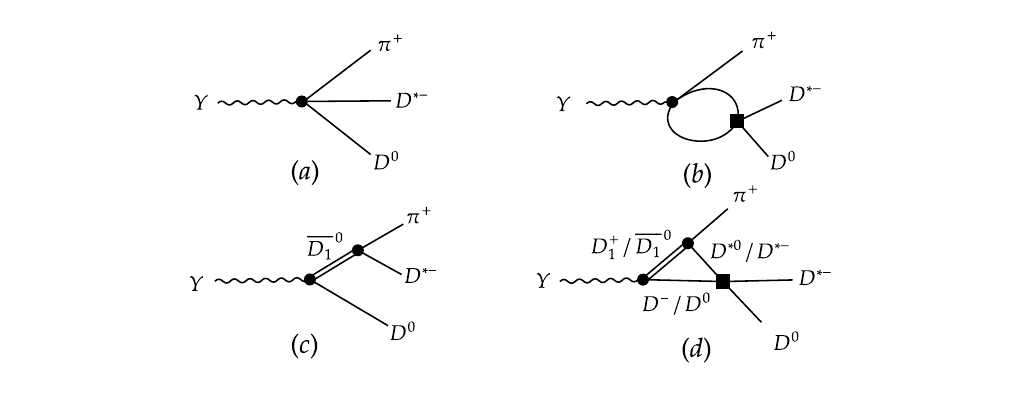}
\caption{The schematic Feynman diagrams for the process $Y(4230/4260)\to \pi^+ D^0 D^{*-}$. The solid circles and boxes denote the interaction vertex and $T$-matrix characterizing the final state interactions. The intermediate channels for bubble loops in (b) and (c) can include $\pi^- J/\psi$ and $[D\bar{D}^*]^-$.}.
\label{fig: feyn diagram DD}
\end{figure*}

\begin{figure*}
    \centering
    \includegraphics[width=0.95\linewidth]{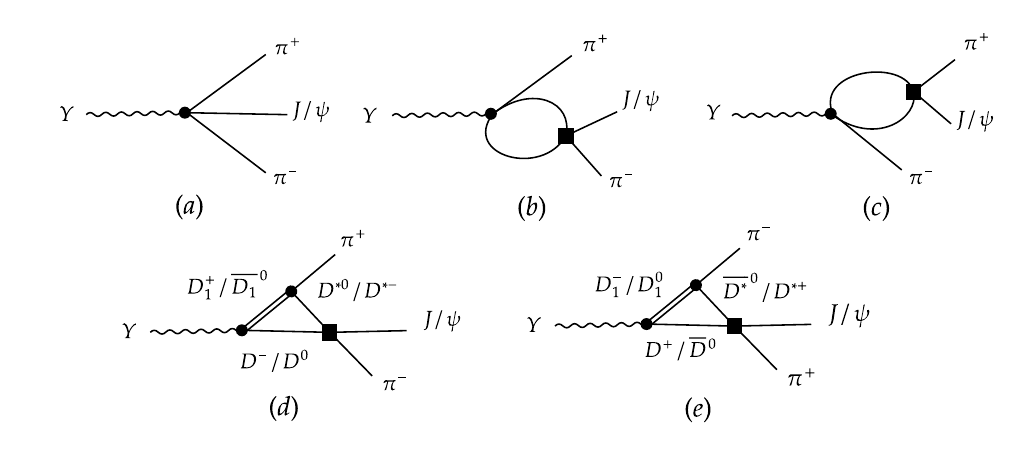}
    \caption{The schematic Feynman diagrams for the process $Y(4230/4260)\to \pi^+\pi^- J/\psi$. The notations are same as in Fig.\ref{fig: feyn diagram DD}. The intermediate channels for bubble loops in (b) and (c) can include $\pi^- J/\psi$ and $[D\bar{D}^*]^-$. $\pi^+$ and $\pi^-$ are treated equivalently due to the isospin symmetry.}
    \label{fig: feyn diagram pijpsi}
\end{figure*}
\begin{figure*}
    \centering
    \includegraphics[width=0.95\linewidth]{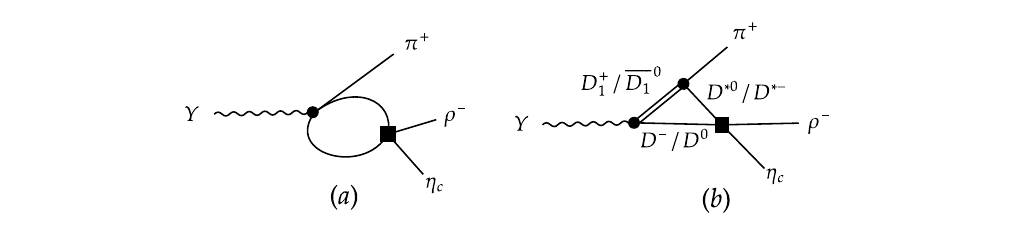}
\caption{The schematic Feynman diagrams for the process $Y(4230/4260)\to \pi^+ \rho^- \eta_c$. The notations are same as in Fig.~\ref{fig: feyn diagram pijpsi}. The intermediate channels for bubble loops in (a) can  include $\pi^- J/\psi$ and $[D\bar{D}^*]^-$. }. 
\label{fig: feyn diagram rhoetac}
\end{figure*}

Since the $Z_c$ structure was discovered in $e^+e^-$ collisions at $\sqrt{s} = 4.23$ and $4.26$ GeV, where the exotic state $Y(4230/4260)$ resides, we investigate the invariant mass distribution through the processes $Y(4230/4260) \to \pi \left(\pi J/\psi, \rho \eta_c, D\bar{D}^*\right)$. For convenience, we denote the $[D\bar{D}^*]^\pm$ states, which have definite positive G-parity, as follows:
\begin{align}
    [D\bar{D}^*]^- = \frac{1}{\sqrt{2}}\left(
    D^0 D^{*-} + D^-D^{*0}   \right),
    \\
    [D\bar{D}^*]^+ = \frac{1}{\sqrt{2}}\left(D^+\bar{D}^{*0} + D^0\bar{D}^{*+}\right),
\end{align}
with the conventions $D \xrightarrow{C} \bar{D}$ and $D^* \xrightarrow{C} -\bar{D}^*$.

The schematic Feynman diagrams for the decay of $Y(4230/4260)$ are illustrated in Figs.~\labelcref{fig: feyn diagram pijpsi,fig: feyn diagram DD,fig: feyn diagram rhoetac} for overview. The black solid box represents the $T$-matrix, which characterizes the final state interactions (FSI) and will be discussed in detail later. 
We have considered the hypothesis that $Y(4230/4260)$ is strongly coupled to a $\bar{D}_1(2420) D$, as suggested in Refs.~\cite{Wang:2013cya, Cleven:2013mka}.
\footnote{For simplicity, we will refer to $D_1(2420)$ and $Y(4230/4260)$ as $D_1$ and $Y$, respectively, unless otherwise specified.}

\subsection{The Lagrangians}

In this subsection, we present the relevant Lagrangians and interaction vertices. The Lagrangians associated with  $Y$ field are given as follows~\cite{Du:2022jjv, Wang:2020axi}:
\begin{align}
    \mathcal{L}_{YD_1D} &=  - g_{YD_1D} \, Y_i D_1^i \bar{D},
    \\
    \mathcal{L}^S_{Y\pi \pi J/\psi} &= c^S_{\pi J/\psi}   \pi^+ \pi^- Y_i \psi^i,
    \\
    \mathcal{L}^D_{Y\pi \pi J/\psi} &= c^D_{\pi J/\psi} Y_i\psi_j  \left[\pi^+ \left(\partial^i\partial^j - \frac{1}{3}\vec{\partial}^2 \delta^{ij} \right) \pi^- \right] \nonumber\\
    &\qquad + \left( \pi^+\leftrightarrow \pi^- \right),
    \\
    \mathcal{L}^S_{Y \pi D\bar{D}^*} &= c^S_{D\bar{D}^*} Y^i \psi_i\left(D^0 D^{*-} + D^-D^{*0}\right)\pi^+,
\end{align}
where $c^L_{\alpha}$ is the coupling constant for the interaction between $Y$ and $\pi+\alpha$ ($\alpha = \pi^- J/\psi,\rho^-\eta_c,D^0 D^{-}$)  with $L$-wave between $\pi$ and $\alpha$-channel denoted by $\pi_L(\alpha)$. Note that for $Y\to\pi D\bar{D}^*$, we only consider the $S$-wave $\pi_S(D\bar{D}^*)$ interaction, as the typical momentum of the final states is not sufficiently large to produce significant $D$-wave contributions. In contrast, for the decay $Y\to\pi\pi J/\psi$, a $D$-wave term $\pi_D(\pi J/\psi)$ is included due to the significantly larger phase space. Here, we neglect the Lagrangian for $Y\to\pi\rho\eta_c$ because experimental data suggest that this vertex is highly suppressed \cite{Yuan:2018inv}. Notably,  we will later demonstrate that the $Y\to\pi\pi J/\psi$ vertex is also suppressed, consistent with the expectations from the heavy quark symmetry. 

The Lagrangian for $D_1D^*\pi$ vertex reads~\cite{Du:2022jjv,Guo:2020oqk} 
\begin{align}
\label{eq:lagD1Dpi}
    \mathcal{L}^S_{D_1 D^*\pi} &= i \frac{2h_S}{\sqrt{3}f_\pi} (D_{1}^{+i} D_0^{*i\dagger} \partial^0 \pi^- ) ,
    \\
    \mathcal{L}^D_{D_1 D^*\pi} &= -\frac{2h_D}{\sqrt{3}f_\pi}\left(D^{+i}_{1} D_0^{*i\dagger} \vec{\partial}^2 \pi^- - 3 D^{+i}_{1}D_0^{*j\dagger} \partial^i\partial^j\pi^- \right)
\end{align}
with $f_\pi=132$ MeV, $\abs{h_S}=0.57$ and $\abs{h_D}=1.17\,\text{GeV}^{-1}$ determined by reproducing the decay widths of $D_2$ and $D_1(2420)$.
%
%
In fact, the contribution from the $D$-wave component would shift the peak of the line shape closer to the threshold, as the energy of the pion increases when the invariant mass of the $D^0 D^{*-}$ system approaches the threshold.
%
This is one of the reason why we have neglected the contribution from the process $e^+ e^- \to D_0^{*}(2300)\bar{D}^* \to (D\pi)\bar{D}^*$ despite the threshold of $ D_0^{*}(2300)\bar{D}^*$ being near 4.23 GeV, since only $S$-wave is allowed for $D_0^*(2300)\to D\pi$ and the peak of its line shape is at around $3.95$ GeV, which is about $100$ MeV away from the experimental peak.

%

%

Finally, we present the Lagrangian for the final state interaction between the three coupled channels.  For the OBE interaction between $D\bar{D}^*$ and $D^*\bar{D}$, the relevant Lagrangians $\mathcal{L}_{D^{(*)}D^{(*)}M}$ with $M=\pi, \rho, \omega$ are given in Refs.~\cite{Li:2012ss,Li:2012cs}. For the values of involved parameters, we follow a recent work~\cite{Wang:2023ovj}, where the line shapes of the double-charmed exotic state $T_{cc}^+$ are successfully reproduced. The interactions between the hidden-charmed channels, i.e., ${\pi J/\psi - \pi J/\psi}, {\rho \eta_c - \rho \eta_c}, {\pi J/\psi - \rho \eta_c}$, are negligible due to the OZI suppression~\cite{Yokokawa:2006td,Liu:2012dv}, which is also confirmed by the HALQCD calculation~\cite{HALQCD:2016ofq}.

For the $D\bar D^*$ coupled with the  hidden-charmed channels, we adopt the effective Lagrangian respecting the heavy quark spin and SU(3) flavor symmetry that reads~\cite{Jenkins:1992nb,Yamaguchi:2019djj},
\begin{align}
    \label{eq:LagHHR}
    \mathcal{L}_{HHR} &= g^\prime \langle R^{(c\bar{c})} \left(\bar{H}^{(\bar{c})}_a \overleftrightarrow{\slashed{\partial}} \bar{H}^{(c)}_a \right)  \rangle + \text{h.c.}  \,, 
\end{align}
with the superfields
\begin{align}
    \bar{H}_a^{(c)} &= \left( {{P_a^*}^\mu}^\dagger \gamma_\mu + P_a^\dagger \gamma_5 \right)\frac{1+\slashed{v}}{2},
    \\
    \bar{H}_a^{(\bar{c})} &= \frac{1-\slashed{v}}{2}\left( \tilde{P_a^*}_\mu^\dagger \gamma^\mu + \tilde{P_a}^\dagger \gamma_5 \right),
    \\
    R^{c\bar{c}} & = \frac{1+\slashed{v}}{2} \left( \psi^\mu \gamma_\mu - \eta_c \gamma_5 \right) \frac{1-\slashed{v}}{2},
\end{align}
where $P = (D^0,D^+,D_s^+)$, $P^* = (D^0,D^+,D_s^+)$, $\tilde{P} = (\bar{D}^0,D^-,D_s^-)$, $\tilde{P^*}=(\bar{D}^{*0},D^{*-},D_s^{*-})$ are SU(3) triplets of the fields that annihilates the corresponding meson.
$\psi^\mu$ and $\eta_c$ are the singlets of field that annihilates corresponding charmonium. 
$v$ is the typical four-velocity of the heavy quark. 
In the heavy quark limit, $v^\mu=(1,\vec{0})$. 
$\langle\cdots\rangle$ denotes the trace in spinor space and $\overleftrightarrow{\partial}\equiv\overrightarrow{\partial}-\overleftarrow{\partial}$. 
After expansion, Eq.~\eqref{eq:LagHHR} can be divided into 
\begin{align}
    \mathcal{L}_{\psi DD^*} &= 2ig^\prime \epsilon^{\mu\nu\alpha\beta} v_\alpha \psi_\beta \left( \tilde{P_{\nu}^*}^\dagger \overleftrightarrow{\partial_\mu}P^\dagger + P_{\nu}^{*\dagger}\overleftrightarrow{\partial_\mu}\tilde{P}^\dagger \right),
    \label{eq:HQETcharomium1}
    \\
    \mathcal{L}_{\psi DD} &= 2g^\prime \psi^\mu P^\dagger \overleftrightarrow{\partial_\mu}\tilde{P}^\dagger ,
    \label{eq:HQETcharomium2}
    \\
    \mathcal{L}_{\psi D^*D^*} &= 2g^\prime\left(\psi^\nu \tilde{P_\nu^*}^\dagger \overleftrightarrow{\partial^\mu} P_\mu^{*\dagger} - \psi^\nu {P_\nu^*} \overleftrightarrow{\partial^\mu} \tilde{P_\mu^*}^\dagger \right.
    \nonumber\\
    &\left.+ \psi_\mu P^{*\nu\dagger} \overleftrightarrow{\partial^\mu} \tilde{P_\nu^*}^\dagger  \right) ,
    \label{eq:HQETcharomium3}
    \\
    \mathcal{L}_{\eta_c D^* D^*} &= -2ig^\prime \epsilon^{\mu\nu\alpha\beta}v_\alpha \eta_c \left(  P_\beta^{*\dagger}\overleftrightarrow{\partial_\mu} \tilde{P_\nu^*}^{\dagger}  \right) ,
    \label{eq:HQETcharomium4}
    \\
    \mathcal{L}_{\eta_c DD^*} &= 2g^\prime \eta_c \left( P_\mu^{*\dagger} \overleftrightarrow{\partial^\mu} \tilde{P}^\dagger  + P^\dagger \overleftrightarrow{\partial^\mu} \tilde{P_\mu^*}^\dagger \right). 
    \label{eq:HQETcharomium5}
\end{align}
The hermitian conjugate terms are omitted for simplicity. 
%

\subsection{The amplitudes}
In this section we present the squared and initial spin-averaged amplitude. 
We denote $s_\alpha \equiv (p_{\alpha_1}+p_{\alpha_2})^2$ as the invariant mass of the $\alpha$-channel from now on. 
Firstly we discuss the $T$-matrix describing the FSI. 
For the present work, all three couple channels are of PV(pseudoscalar-vector) type so the $T$-matrix is given by the following couple-channels Lippmann-Schwinger Equation(LSE),
\begin{align}\label{eq:LSE}
&T_{\alpha\beta}(\boldsymbol{p},\sigma_{\alpha_1},\boldsymbol{k},\sigma_{\beta_1};E) = V_{\alpha\beta}(\boldsymbol{p},\sigma_{\alpha_1},\boldsymbol{k},\sigma_{\beta_1}) 
+ \sum\limits_{\sigma_{\gamma_1},\gamma} \int d^3\boldsymbol{q}
\nonumber\\ 
&\,\,\,
\times V_{\alpha\gamma}(\boldsymbol{p},\sigma_{\alpha_1},\boldsymbol{q},\sigma_{\gamma_1})
G_{\gamma}(\boldsymbol{q};E) T_{\gamma,\beta}
(\boldsymbol{q},\sigma_{\gamma_1},\boldsymbol{k},\sigma_{\beta_1};E),
\end{align}
where $\sigma_{\alpha_1}$ is the polarization indices of vector meson in the $\alpha$-channel and $G_{\alpha}(\boldsymbol{q},E)$ is the propagator for intermediate $\alpha$-channel given by
\footnote{The width of $\rho$-meson are neglected here. Since the potential $V_{D\bar{D}^*-\rho\eta_c}$ vanishes at the $\rho\eta_c$ threshold, it will not produce a cusp at the  $\rho\eta_c$ threshold which is not observed. The finite width of $\rho$ hardly affect the result of this work.}

\begin{align}
    G_{\alpha}(\boldsymbol{q},E) &= \frac{1}{(2\pi)^3 4\omega_{\alpha_1}(|\boldsymbol{q}|)\omega_{\alpha_2}(|\boldsymbol{q}|) }\nonumber\\
    &\times\frac{1}{\left(E-\omega_{\alpha_1}(|\boldsymbol{q}|)-\omega_{\alpha_2}(|\boldsymbol{q}|)+i0^+\right)  },
    \label{eq:G}
\end{align}
where $\omega_{\alpha_i}(q)=\sqrt{q^2+m_{\alpha_i}^2}$. Practically, LSE is solved in the partial-wave representation where the integral equation is reduced to one-dimensional form,
\begin{align}\label{eq:partialwaveLSE}
    &T^{J }_{l_\beta,l_\alpha}(p,k;E) = V^J_{l_\beta, l_\alpha}(p,k)
    \notag\\
    &+ \sum\limits_{l_\gamma} \int q^2 dq V^J_{l_\beta,l_\gamma}(p,q)G_\gamma(q;E)T^J_{l_\gamma,l_\alpha}(q,k;E),
\end{align}
with 
\begin{align*}
    V^J_{l^\prime,l}(p,k)& = \sum\limits_{\sigma \sigma^\prime} \int d \Omega_{\boldsymbol{p}}\sqrt{\frac{4\pi(2l + 1)}{2J+1}} C_{l^\prime 1}(J\sigma ;\sigma-\sigma^\prime,\sigma^\prime)
    \nonumber\\
    &\times
    C_{l1}(J\sigma,0\sigma) Y_{l^\prime,\sigma-\sigma^\prime}(\hat{\boldsymbol{p}}) V(\boldsymbol{p},\sigma^\prime,k\boldsymbol{\hat{e_z}},\sigma), 
\end{align*}
where $C_{j_1j_2}(JM;m_1m_2)$ and $Y_{lm}$ are the Clebsch-Gordon coefficients and spherical harmonics, respectively. 
Based on the partial wave analysis by BES\uppercase\expandafter{\romannumeral3}~\cite{BES3pj423}, we only focus on $J=1$ and $l=l^\prime=0$ in the present paper. 
The numerical method to solve the partial-wave LSE is given in Appendix~\ref{append:LSE}.
The kernel of LSE, $V$, is related to the tree-level Lorentz-invariant amplitude $i\mathcal{M}$ by 
\begin{align}
    V = -\mathcal{M} .
\end{align}

It is worth mentioning several subtle issues here. 
Firstly, we note that all the heavy-meson and charmonium fields in Eqs.~\eqref{eq:lagD1Dpi} to \eqref{eq:HQETcharomium5} are not normalized as in the relativistic field theory, where the particles are normalized as $\braket{\vec{p}^\prime}{\vec{p}} = (2\pi)^3 2 E_{\vec{p}}\, \delta^3(\vec{p}^\prime-\vec{p})$. They differ from a factor $\sqrt{m}$ at the leading order. Therefore, to obtain $i\mathcal{M}$, extra factors $\sqrt{m}$ with $m$ the mass of heavy-meson or charmonium should be multiplied when evaluating the Feynman diagrams with the lagrangian above. 
Secondly, since a factor $m_c v^\mu$ has been already subtracted when effective Lagrangians in Eqs.(\ref{eq:lagD1Dpi}-\ref{eq:HQETcharomium5}) are constructed, the differential operator acting on the heavy-meson field gives the residual momentum $k^\mu := p^\mu - m_c v^\mu$ where $p$ is the four momentum. 
Thirdly, in the LSE or rescattering equation in the Time-Ordered-Perturbation-Theory formalism, only the conservation of three-momentum are promised at each vertex. Therefore, there are at least two schemes to evaluate the transferred momentum $q$ in $\mathcal{M}$ which make differences unless the energy-conservation condition are imposed.
\begin{itemize}
    \item Scheme 1~\cite{Blankenbecler:1965gx}
    : \begin{align}
        V_{\beta\alpha} = -\mathcal{M}\left(q_0 = \frac{1}{2}\left( \omega_{\alpha_1} - \omega_{\beta_1} + \omega_{\beta_2} - \omega_{\alpha_2}  \right)  \right).  
    \end{align} 
    \item Scheme 2~\cite{Wu:2012md}: \begin{align}
        V_{\beta\alpha} = - \frac{1}{2}\left(
    \mathcal{M}\left(q_0=\omega_{\alpha_1} - \omega_{\beta_1}\right) + \mathcal{M}\left(q_0=\omega_{\beta_2} - \omega_{\alpha_2}\right) 
    \right).
    \end{align}
\end{itemize}
For scheme 1, the potential would degenerate to a widely-used form $V\propto \left(\boldsymbol{q}^2 + m_\text{eff}^2 + i\epsilon\right)^{-1}$ if $\omega_{\alpha(\beta)_i}\approx m_{\alpha(\beta)_i}$ is further applied. 
Both schemes are used later and the scheme-dependence of result are investigated.
Lastly, to eliminate the ultraviolet divergence of the loop integral in the LSE that arising from the short distance physics, a non-local dipole form factor is introduced
\begin{align}
    V_{\beta\alpha}(\boldsymbol{p},\boldsymbol{k}) \to V_{\beta\alpha}(\boldsymbol{p},\boldsymbol{k})\left(\frac{\Lambda_\alpha^2}{ \Lambda_\alpha^2 + \boldsymbol{k}^2}\right)^2 \left(\frac{\Lambda_\beta^2}{\Lambda_\beta^2+\boldsymbol{p}^2}\right)^2.
\end{align}
The $\Lambda_{[D\bar{D}^*]}$ is fixed at 1 GeV following Ref.~\cite{Wang:2023ovj}, and it has been shown that its variation within a reasonable range can be effectively absorbed into coupling constant.
Given that $\frac{m_{J/\psi}}{m_D}\approx \frac{m_{\eta_c}}{m_D} \approx 1.5$, the cutoff $\Lambda_{\pi J/\psi} = \Lambda_{\rho \eta_c}$ is chosen to be fixed at $1.5$ GeV. Similarly, as will be shown later, within a reasonable range of 
 $\Lambda_{\rho\eta_c,\pi J/\psi}$, its influence can be effectively absorbed into the coupling constant $g^\prime$.

Using the $T$-matrix,we can now explicitly express $\abs{\mathcal{M}_\alpha}^2 = \abs{\mathcal{M}_\alpha^S}^2 + \abs{\mathcal{M}_\alpha^D}^2 $ as follows,
\begin{widetext}
\begin{align}
   \abs{\mathcal{M}^S_{D^0D^{*-}}}^2 &= \abs{ 
   c^S_{D\bar{D}^*} + \sum\limits_{\alpha=\pi J\psi,[D\bar{D}^*]} c^S_\alpha \mathcal{I}_{\alpha\to [D\bar{D}^*]} + g_{YD_1D}h_S^\prime \omega_\pi \left( \frac{1}{s_{D^*\pi}-m_{D_1}^2+im_{D_1}\Gamma_{D_1}} + \mathcal{Q}_{D\bar{D}^*}  \right)
   }^2,
   \\
   \abs{\mathcal{M}^D_{D^0D^{*-}}}^2 &= \frac{2}{9}p_\pi^4 \abs{ 
    c^D_{\pi J/\psi} \mathcal{I}_{\pi J/\psi \to [D\bar{D}^*]} + g_{YD_1D}h_D^\prime  \left( \frac{1}{s_{D^*\pi}-m_{D_1}^2+im_{D_1}\Gamma_{D_1}} + \mathcal{Q}_{D\bar{D}^*} \right)
   }^2,
   \\
   \abs{\mathcal{M}^S_{\pi^- J/\psi}}^2 &= 
    \left|c^S_{\pi J/\psi} + \sum\limits_{\alpha=\pi J/\psi,[D\bar{D}^*]} x_\alpha c^S_\alpha \left( \mathcal{I}_{\alpha\to\pi^+ J/\psi} + \mathcal{I}_{\alpha\to\pi^- J/\psi}\right) 
    + \mathcal{F}^+ +\mathcal{F}^- \right|^2,
   \\
   \abs{\mathcal{M}^D_{\pi^- J/\psi}}^2 &= 
   \frac{2}{9}p_{\pi^+}^4\abs{\mathcal{A^+}}^2 + 
   \frac{2}{9}p_{\pi^-}^4\abs{\mathcal{A^-}}^2 +
   \frac{2}{9} p_{\pi^+}^2 p_{\pi^-}^2 \left(3\cos^2\theta_{\pi^+\pi^-} - 1 \right)\Re\left( \mathcal{A}^{+*} \mathcal{A}^-   \right),
   \\
   \abs{\mathcal{M}^S_{\rho\eta_c}}^2 &= \abs{ 
   \sum\limits_{\alpha=\pi J/\psi,[D\bar{D}^*]}x_\alpha c_\alpha^S\mathcal{I}_{\alpha\to\rho\eta_c} + \sqrt{2} g_{YD_1D} h^\prime_S\mathcal{Q}_{\rho\eta_c}
   }^2 ,
   \\
   \abs{\mathcal{M}^D_{\rho\eta_c}}^2 &= \frac{2}{9}p_\pi^4\abs{ 
    c_{\pi J/\psi}^D\mathcal{I}_{\alpha\to\rho\eta_c} + \sqrt{2} g_{YD_1D} h^\prime_D\mathcal{Q}_{\rho\eta_c}
   }^2 ,
\end{align}
where,
\begin{align}
   \mathcal{A}^\pm &= c^D_{\pi J/\psi} +  c^D_{\pi J/\psi} \mathcal{I}_{\pi^\mp J/\psi \to \pi^\mp J/\psi}  + \sqrt{2} g_{YD_1D} h^\prime_D \mathcal{Q}_{\pi^\mp J/\psi},\\
   \mathcal{F}^\pm &= \sqrt{2} g_{YD_1D}h_S^\prime \omega_\pi(p_{\pi^{\pm}})\mathcal{Q}_{\pi^{\mp}J/\psi}  ,
\end{align}
\end{widetext}
and  $h_{S}^\prime = \frac{\sqrt{4m_{D_1}m_{D^*}}}{\sqrt{3}f_\pi}h_{S}$, 
$h_{D}^\prime = -\frac{2\sqrt{3m_{D_1}m_{D^*}}}
{f_\pi}h_{D}$, $x_{\pi J/\psi([D\bar{D}^*])}=1(\sqrt{2})$. 
Additionally, 
$\omega_\pi = \sqrt{p_\pi^2+m_\pi^2}$ and $p_\pi = \lambda^\frac{1}{2}\left(\sqrt{s},\sqrt{s_\alpha},m_\pi\right)$ with $\lambda(a,b,c)$ being the modified Källén triangle function,
\begin{align}
   \lambda(a,b,c) \equiv \frac{\left(a^2-(b+c)^2\right)\left(a^2-(b-c)^2\right)}{4a^2} .
\end{align}
The relative angle $\cos\theta_{\pi^+\pi^-}$ is given by 
\begin{align}
    \cos\theta_{\pi^+\pi^-} &= \frac{ s_{\pi^+\pi^-} -2m_\pi^2 - 2\omega_{\pi^-}\omega_{\pi^+}}{2p_{\pi^+}p_{\pi^-}},
    \\
    s_{\pi^+\pi^-} &= s + 2m_\pi^2 + m_{J/\psi}^2 - s_{\pi^+ J/\psi} - s_{\pi^- J/\psi}
\end{align}
$\mathcal{I}_\alpha$ and $\mathcal{Q}_\alpha$ comes from the FSI in the bubble diagram and triangle diagram, respectively. The explicit expressions are given by
\begin{align}\label{eq:bubble}
 &    \mathcal{I}_{\alpha\to\beta} = \nonumber \\
    &\int \frac{q^2 dq}{(2\pi)^3 4\omega_{\alpha_1}(q)\omega_{\alpha_2}(q)}\frac{T^{J=1}_{0,0;\alpha\to\beta}(\bar{p}_\beta,q;\sqrt{s_\beta})}{\sqrt{s_\beta} - \omega_{\alpha_1}(q) - \omega_{\alpha_2}(q) + i0^+},
\end{align}
and 
\begin{align}\label{eq:triangle}
    \mathcal{Q}_\alpha = \int \frac{d^3\boldsymbol{q}}{(2\pi)^3 4\pi}\frac{T^{J=1}_{0,0;D\bar{D}^*\to\alpha}(q^*;\bar{p}_\alpha;\sqrt{s_\alpha})}{8\omega_{D_1}(\abs{\boldsymbol{p_\pi}+\boldsymbol{q}})\omega_{D^*}(q)\omega_{D}(\abs{\boldsymbol{p_\pi}+\boldsymbol{q}})}
    \notag\\\times
    \frac{1}{\left(\sqrt{s}-\omega_{D_1}(\abs{\boldsymbol{p_\pi}+\boldsymbol{q}})-\omega_{D}(\abs{\boldsymbol{p_\pi}+\boldsymbol{q}}) + i\frac{\Gamma_{D_1}}{2}\right)}
    \notag\\\times
    \frac{1}{\left(\sqrt{s}-\omega_\pi(p_\pi)-\omega_{D^*}(q) - \omega_{D}(\abs{\boldsymbol{p_\pi}+\boldsymbol{q}}) + i0^+  \right)},
\end{align}
where $\bar{p}_\beta \equiv 
\lambda^{\frac{1}{2}}\left(\sqrt{s_\beta},m_{\beta_1},m_{\beta_2}\right)$ is the on-shell momentum in the rest frame of $\beta$-channel. 
The integration variable in Eq.~(\ref{eq:triangle}) is the momentum of $D^*$ in the rest frame of $Y$ while $q^*$ is that in the rest frame of $D\bar{D}^*$ and hence given by
$q^*=\sqrt{\left(\gamma\sqrt{q^2+m_{D^*}^2} + \gamma\beta q \cos\theta\right)^2-m_{D^*}^2}$ with 
$\gamma = \sqrt{\frac{p_\pi^2+s_\alpha}{s_\alpha}}$ and $\gamma\beta = -\frac{p_\pi}{\sqrt{s_\alpha}}$. 
Note that the finite width of $\Gamma_{D_1(2420)}=31.3$ MeV has been taken into account. 
The numerical method for the calculation of $\mathcal{Q}_\alpha$ is discussed in Appendix.~\ref{append:triangle}.

As the final result, the invariant mass distribution of $Y\to\pi\alpha$ reads
\begin{align}
    \frac{d\Gamma_{Y\to\pi^+\alpha}}{d\sqrt{s_\alpha}} = \mathcal{N}_\alpha \int\limits_{t^-_\alpha(s_\alpha)}^{t^+_\alpha(s_\alpha)} \frac{\abs{\mathcal{M_\alpha}}^2}{(2\pi)^3 16 \sqrt{s^3} }\sqrt{s_\alpha} \, d t_\alpha  + \mathcal{B_\alpha},
\end{align}
with $t_{\pi^-J/\psi}=s_{\pi^+J/\psi},t_{\rho^-\eta_c}=s_{\rho^-\pi^+}$ and $t_{D^0D^{*-}}=s_{\pi^+ D^{*-}}$. 
$t_\alpha^{\pm}$ is the upper and lower limit of Mandelstam variables~\cite{Workman:2022ynf}. $\mathcal{N_\alpha}$ are scale factors.
Here, we introduce two polynomial distribution functions $\mathcal{B}_\alpha$ incoherently for $\alpha = \pi J/\psi$ and $D^0D^{*-}$, which is defined by 
\begin{align}
    \mathcal{B}_\alpha = b^\alpha_0 \left(\sqrt{s_\alpha}-m_{\alpha}^-\right)^{b^\alpha_1}\left(m^+-\sqrt{s_{\alpha}}\right)^{b^\alpha_2},
\end{align}
with $m_\alpha^-=m_{\alpha_1}+m_{\alpha_2}$ and $m^+ = \sqrt{s}-m_\pi$. 
For $D\bar{D}^*$, it is introduced to mimic the possible contributions from processes such as $e^+ e^- \to Y(4160) \to \pi D\bar{D}^*$, which may be important to reproduce the tail of line shape~\cite{vonDetten:2024eie}. 
For $\pi J/\psi$, it is introduced to approximate the $\pi\pi$ FSI, which are not explicitly included here as in Ref.~\cite{Du:2022jjv}. 
In a recent work~\cite{Chen:2023def}, the $\pi\pi$ (and even $K\bar{K}$) FSI is carefully treated using dispersion relation to accurately calculate the pole position of $Z_c$. 
Nevertheless, the result do not differ significantly from their previous result in Ref.~\cite{Du:2022jjv}, where the $\pi\pi$ FSI is also mimicked by a polynomial function. 
Therefore, we believe that the treatment here will not significantly impact the conclusions. 
Later, we will see that the contribution from polynomial function has little interference with the  $\abs{\mathcal{M_\alpha}}^2$ term, which justifies their incoherent addition. 
Besides, to reduce the number of fitting parameters, we set $b^{\pi J/\psi}_1=b^{\pi J/\psi}_2$. 

\subsection{parameters in the formalism}

In this subsection, we summarize the free parameters in our formalism. For the $T$-matrix, there is only one free parameter $g^\prime$ as defined in Eq.~(\ref{eq:LagHHR}).
For the vertices involving a $Y$-particle, the free parameters are $g_{YD_1D}$ and $c_\alpha^{S,D}$. 
For the final invariant mass distribution,the free parameters are $N_\alpha$ and $b_{0,1,2}^\alpha$. 
With the exception of $g^\prime$, all parameters are $\sqrt{s}$-dependent, which means they may vary when fitting experimental data at $\sqrt{s}=4.23$ and $4.26$ GeV.
In a word, our model incorporates a total of $2\times 11 +1 = 23$ free parameters to fit over 250 experimental data points at $\sqrt{s}=4.23$ and $4.26$ GeV. 
The number of parameters is somewhat reduced compared to the previous work.

\section{Fitting Results and discussions}
\label{sec:fitting}

\begin{figure*}
    \centering
    \includegraphics[width=0.48\linewidth]{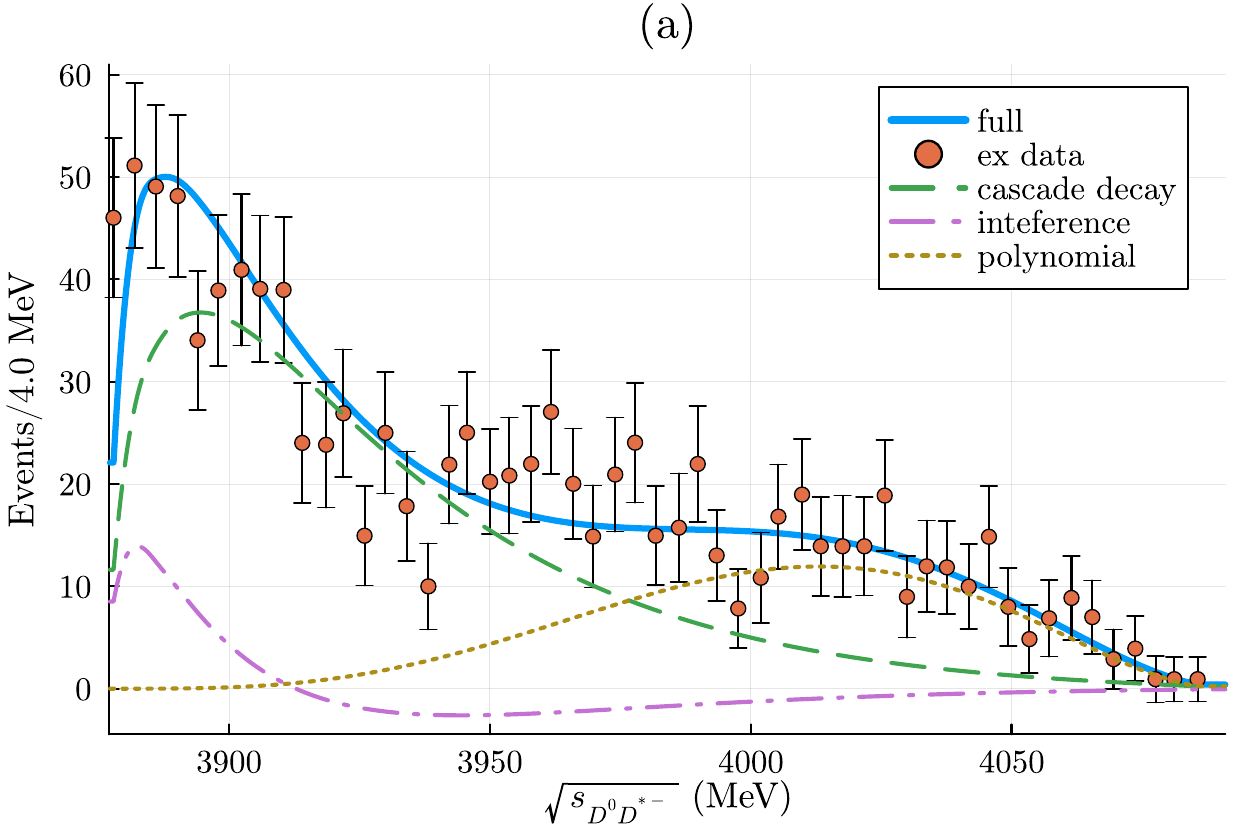}
    \includegraphics[width=0.48\linewidth]{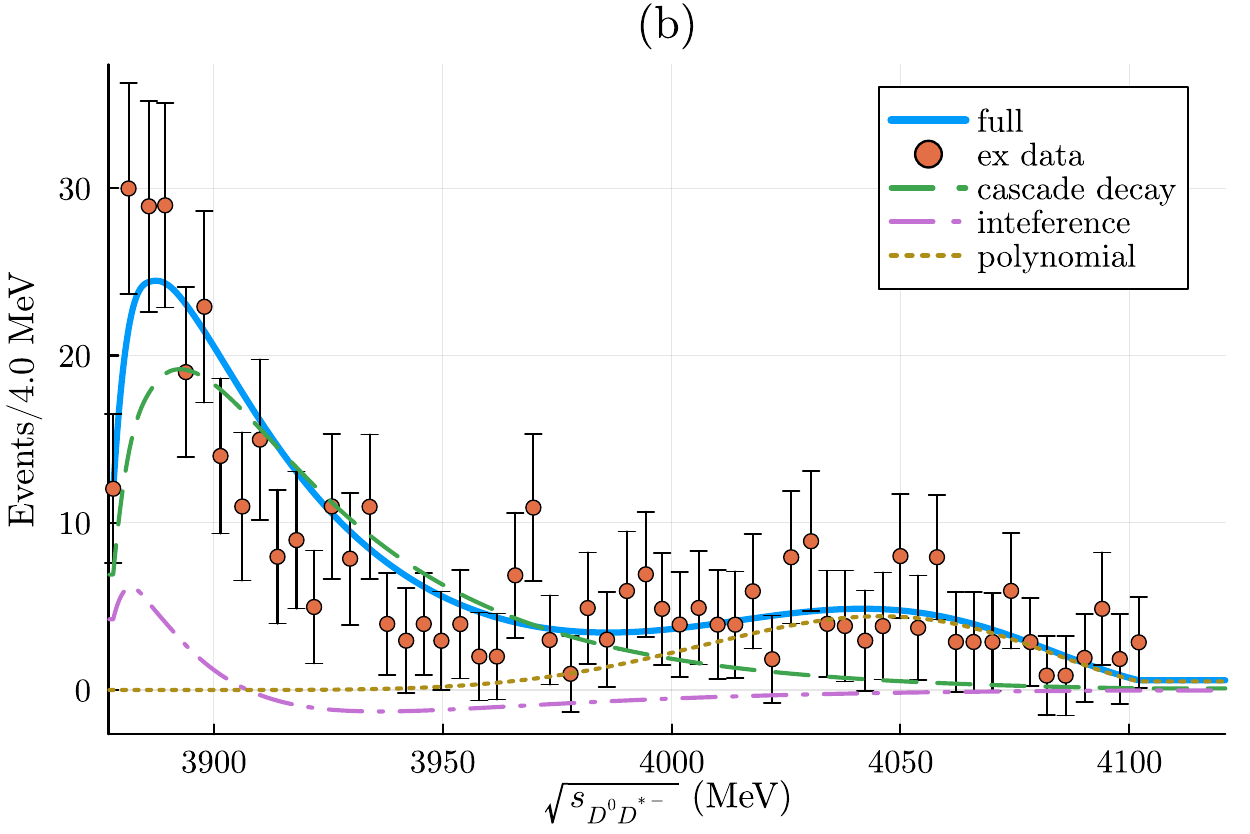}
    \caption{Experimental data as well as theoretical line shapes for $D^0 D^{*-}$ distribution of $e^+ e^- \to \pi^+ D^0 D^{*-}$ process at $(a)\,\sqrt{s}=4.23$ and $(b)\,\sqrt{s}=4.26$ GeV. Orange solid circles are experimental data. Blue solid, green dashed, pink dot-dashed, yellow dotted lines denote the contribution cascade decay, interference between cascade decay and triangle diagram, and the polynomial function, respectively.}
    \label{fig: fitting lineshape 0}
    \includegraphics[width=0.48\linewidth]{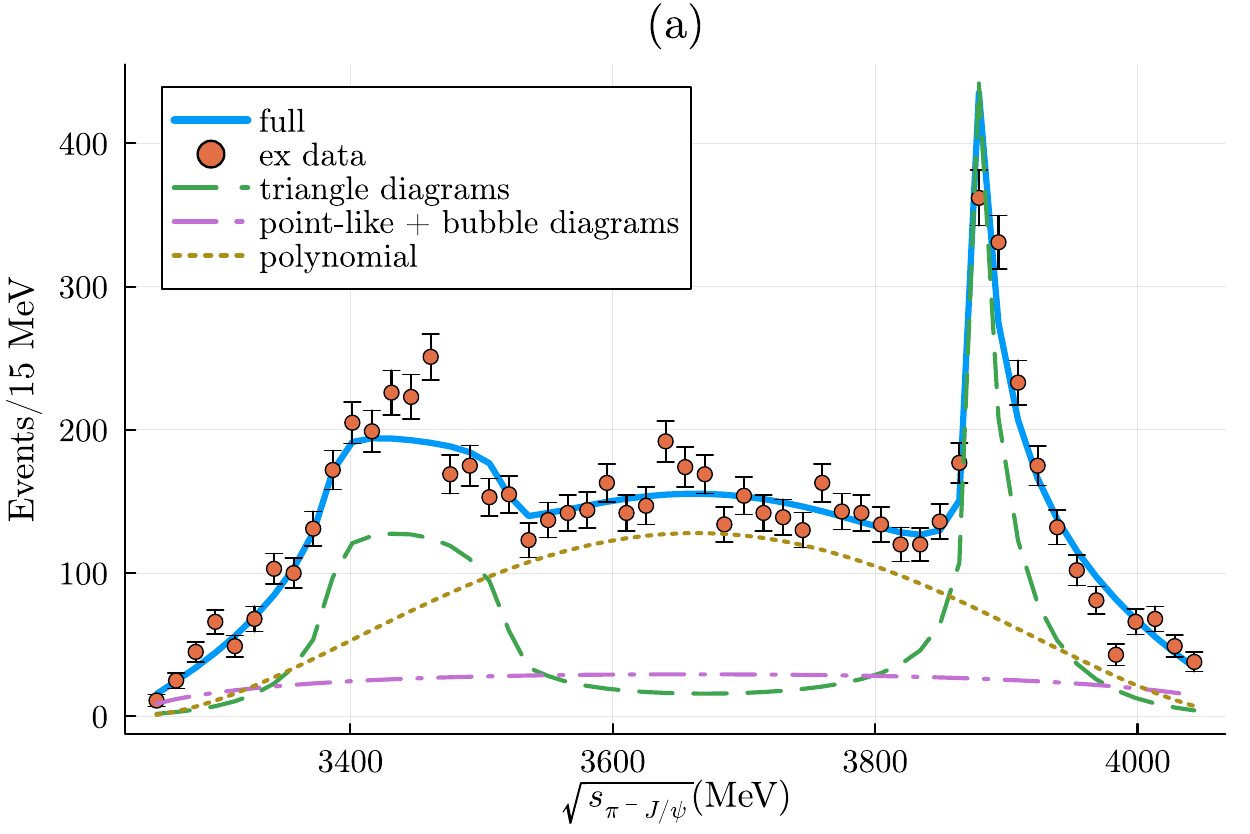}
    \includegraphics[width=0.48\linewidth]{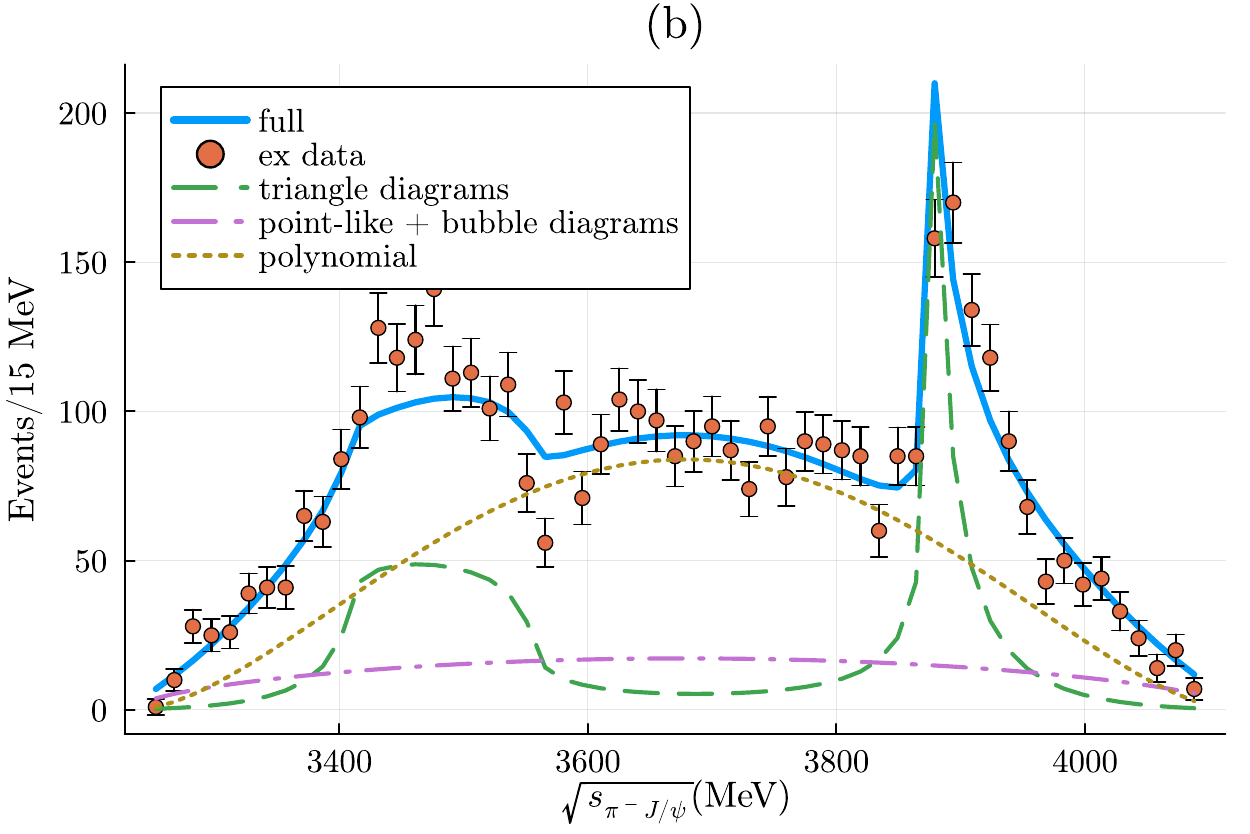}
    \caption{Experimental data as well as theoretical line shapes for $\pi^- J/\psi$ distribution of $e^+ e^- \to \pi^+ \pi^- J/\psi$ process at $(a)\,\sqrt{s}=4.23$ and $(b)\,\sqrt{s}=4.26$ GeV. Orange solid circles are experimental data. Blue solid, green dashed, pink dot-dashed, yellow dotted lines denote the contribution from full amplitude, triangle diagram, point-like as well as bubble diagrams, and the polynomial function, respectively.
    }
    \label{fig: fitting lineshape 1}
\end{figure*}

\begin{figure}
    \centering
    \includegraphics[width=0.95\linewidth]{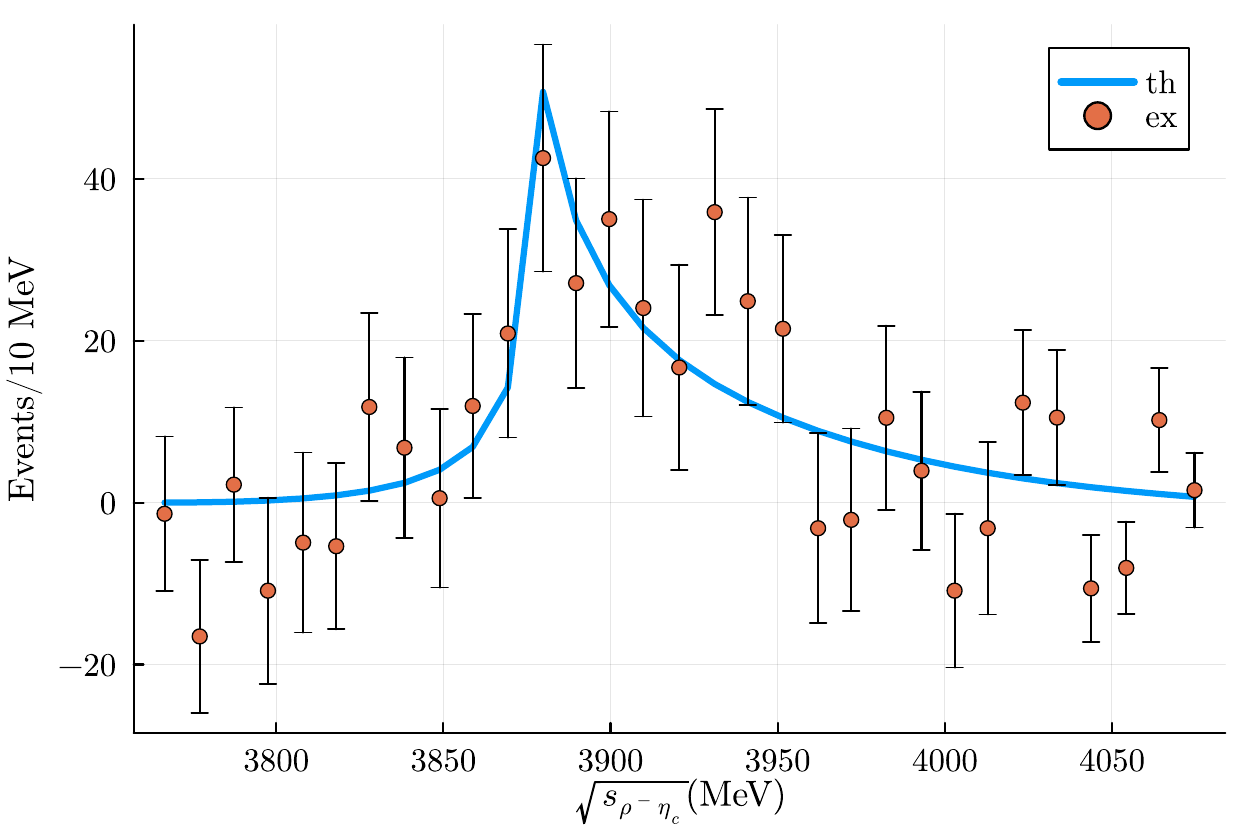}
    \caption{
    Experimental data as well as theoretical line shape for $\rho^-\eta_c$ invariant mass distribution of $e^+e^-\to \pi^+\rho^-\eta_c\to \eta_c \pi^0\pi^+\pi^-$ process at $\sqrt{s}=4.23$ GeV. Orange solid circle and blue solid line denote the experimental data and theoretical line shape, respectively. 
    }
    \label{fig: fitting lineshape 2}
\end{figure}

We are now ready to fit the invariant mass distributions of $\rho^-\eta_c$ at $\sqrt{s}=4.23$ GeV and $\pi^- J/\psi, D^0D^{*-}$ at $\sqrt{s}=4.23$ as well as $4.26$ GeV. The results for $g^\prime$ across two schemes at several fixed cutoffs are presented in Table~\ref{tab: parameters fit}, with values from other references included for comparison. The experimental line shapes are successfully reproduced in both schemes, yielding a reduced chi-squared $\hat{\chi}^2\sim 1.6$. For illustration, we present the results for scheme 1 at $\Lambda = 1.5$ GeV in Figs.~\labelcref{fig: fitting lineshape 0,fig: fitting lineshape 1,fig: fitting lineshape 2}. Specific contributions are identified and depicted with dashed lines for further analysis. 
\subsection{$D\bar{D}^*$ invariant mass distribution}
For the $D\bar{D}^*$ invariant mass distribution, the cascade decay $Y \to \bar{D}_1 D \to D\bar{D}^* \pi$ Fig.~\ref{fig: feyn diagram DD}(c) is found to be the dominant contribution. This can be explained by noting that, because the OBE isovector couple-channel interactions are weak, the contributions from the two rescattering diagrams shown in Fig.~\ref{fig: feyn diagram DD}(b,d) are much smaller compared to the corresponding tree diagrams  Fig.~\ref{fig: feyn diagram DD}(a,c).
%
Nonetheless, the interference between the triangle diagram Fig.~\ref{fig: feyn diagram DD}(d) and the cascade decay Fig.~\ref{fig: feyn diagram DD}(c) is crucial in bridging the gap between the cascade decay and the experimental data at threshold, as illustrated by the purple dot-dashed lines in Fig.\ref{fig: fitting lineshape 0}. The incoherent polynomial function contributes significantly only at the tail, as anticipated.  The point-like tree diagram, depicted in Fig.\ref{fig: feyn diagram DD} (a) and can only produce a smooth line shape.,  contributes very little here and hence not shown in Fig.~\ref{fig: fitting lineshape 0}.
In fact, if the $Y \to \pi_D(D\bar{D}^*)$ interaction were introduced, the contribution of that point-like diagram could be large enough to reproduce the peak even without the cascade decay mechanism. However, this interpretation is implausible, as it would be unnatural for the contribution from $Y \to \pi_D(D\bar{D}^*)$ to exceed that from $Y \to \pi_S(D\bar{D}^*)$ near the threshold. Further details can be referred to Appendix~\ref{append: no tri}. 

Additionally, the bubble diagram $Y \to \pi(\pi J/\psi) \to \pi(D\bar{D}^*)$, as shown in Fig.~\ref{fig: feyn diagram DD}(b,c), could have a sizable contribution if the $Y \to \pi(\pi J/\psi)$ coupling is sufficiently large so that such contribution is comparable to the cascade decay. However, if this were the case, the line shape in the $\pi J/\psi$ distribution would not be accurately reproduced. This shows that the key advantage of our approach is the ability to achieve a correct description by analyzing both datasets together. We also note that in Ref.~\cite{He:2017lhy}, the near-threshold peak appears to be reproduced using the OBE potential without the cascade decay. However, we argue that the formalism in that work may be inconsistent, as in their equation (13) and (14), the tree diagram Fig.~\ref{fig: feyn diagram DD}(a) is artificially suppressed while the rescattering diagram Fig.~\ref{fig: feyn diagram DD}(b) remains unchanged so the FSI is no longer week relatively.

In conclusion, the sharp near-threshold peak in the $D\bar{D}^*$ contribution is produced by the cascade decay and the triangle diagram in our model, consistent with the Refs.~\cite{Wang:2020axi, vonDetten:2024eie, Guo:2014iya, Guo:2019twa}.

We now study the triangle diagram Fig.~\ref{fig: feyn diagram DD}(d) in more detail, which enhance the peak structure observed in the $D\bar{D}^*$ distribution through the interference with cascade decay as just discussed. The enhancement may originate from two sources:  $T$-matrix  in Fig.~\ref{fig: feyn diagram DD} and the kinematic triangle singularity (TS). In Refs.~ \cite{Wang:2013cya, Liu:2013vfa, Guo:2019twa}, it is suggested that the TS could be crucial for the peak structure in the $D\bar{D}^*$ distribution even the system does not fall exactly within the region where the TS occurs, as the $D_1\bar{D}$ threshold is above $4.26$ GeV. This raises the question of whether the enhancement is entirely due to TS or partially due to other factors such as the $T$-matrix.

To investigate this, we replace the $T$-matrix with a constant in Eq.~\eqref{eq:triangle} and present the line shapes produced by the pure triangle integral at $\sqrt{s} = 4.23$ and $4.26$ GeV in Fig.~\ref{fig:puretriangle}(a,b). While the peak near the threshold is evident, it is not sharp enough to reproduce the purple dot-dashed lines in Fig.~\ref{fig: fitting lineshape 0} when interference with the cascade decay is considered.
This indicates that, although the kinematic triangle loop is significant, a $T$-matrix describing the FSI is still necessary in our model. In summary,  the threshold peak in the $D\bar{D}^*$ distribution arises from the combination of $Y \to D\bar{D}_1 \to D\bar{D}^* \pi$, the $T$-matrix for $D\bar{D}^* \to D\bar{D}^*$, and the kinematic triangle loop. The peak cannot be reproduced as sharply without all of these elements.

Parenthetically, an asymmetry parameter $\mathcal{A} = \frac{n_{>0.5} - n_{<0.5}}{n_{>0.5} + n_{<0.5}}$, where $n_{>0.5(<0.5)}$ represents the number of events with $\abs{\cos\theta_{\pi D}} > 0.5(<0.5)$, is defined to study the role of $Y \to \bar{D}_1 D \to D\bar{D}^*\pi$ in the context of $Z_c$ on the experimental side~\cite{BES3DD426}. In Ref.~\cite{BES3DD426}, $\mathcal{A}_{\text{MC}} = 0.43 \pm 0.04$ and $\mathcal{A}_{\text{data}} = 0.12 \pm 0.06$ were obtained using Monte Carlo simulation and experimental data analysis, respectively. This led to the conclusion that $D_1\bar{D}$ contributes only minimally.
However, in Ref.~\cite{Wang:2020axi}, the authors calculate the asymmetry parameter $\mathcal{A}_{\text{th}}$ for the cascade-decay-dominant mechanism of $Z_c$, finding $\mathcal{A}_{\text{th}} \approx 0.13$, which is consistent with the experimental value $\mathcal{A}_{\text{data}}$. Therefore, the interpretation here cannot be ruled out based on the asymmetry parameter $\mathcal{A}$.

\subsection{$\pi^- J/\psi$ and $\eta_c\rho$ invariant mass distribution}

We now turn to the discussion of the fitting results for the $\pi^- J/\psi$ distribution. First, the polynomial function produces a line shape similar to those generated by the $\pi\pi$ FSI in Ref.~\cite{BES3pj423}, as expected. Second, the peak is primarily contributed by the triangle diagram.
To explain why the schematic Feynman diagrams (a,b,c) in Fig.~\ref{fig: feyn diagram pijpsi} are suppressed, we point out that they can only produce a peak if the $Y \to \pi(D\bar{D}^*)$ interaction is sufficiently strong, such that the bubble diagram $Y \to \pi(D\bar{D}^*) \to \pi(\pi J/\psi)$ becomes comparable with the tree diagram $Y \to \pi\pi J/\psi$. However, this level of strength of $Y \to \pi D\bar{D}^*$ is not supported by the experimental data of the $D\bar{D}^*$ distribution.
The discussions on both the $\pi J/\psi$ and $D\bar{D}^*$ distributions demonstrate that the model becomes more constrained when experimental data from different channels are considered.

\begin{figure}
\centering
\includegraphics[width=\linewidth]{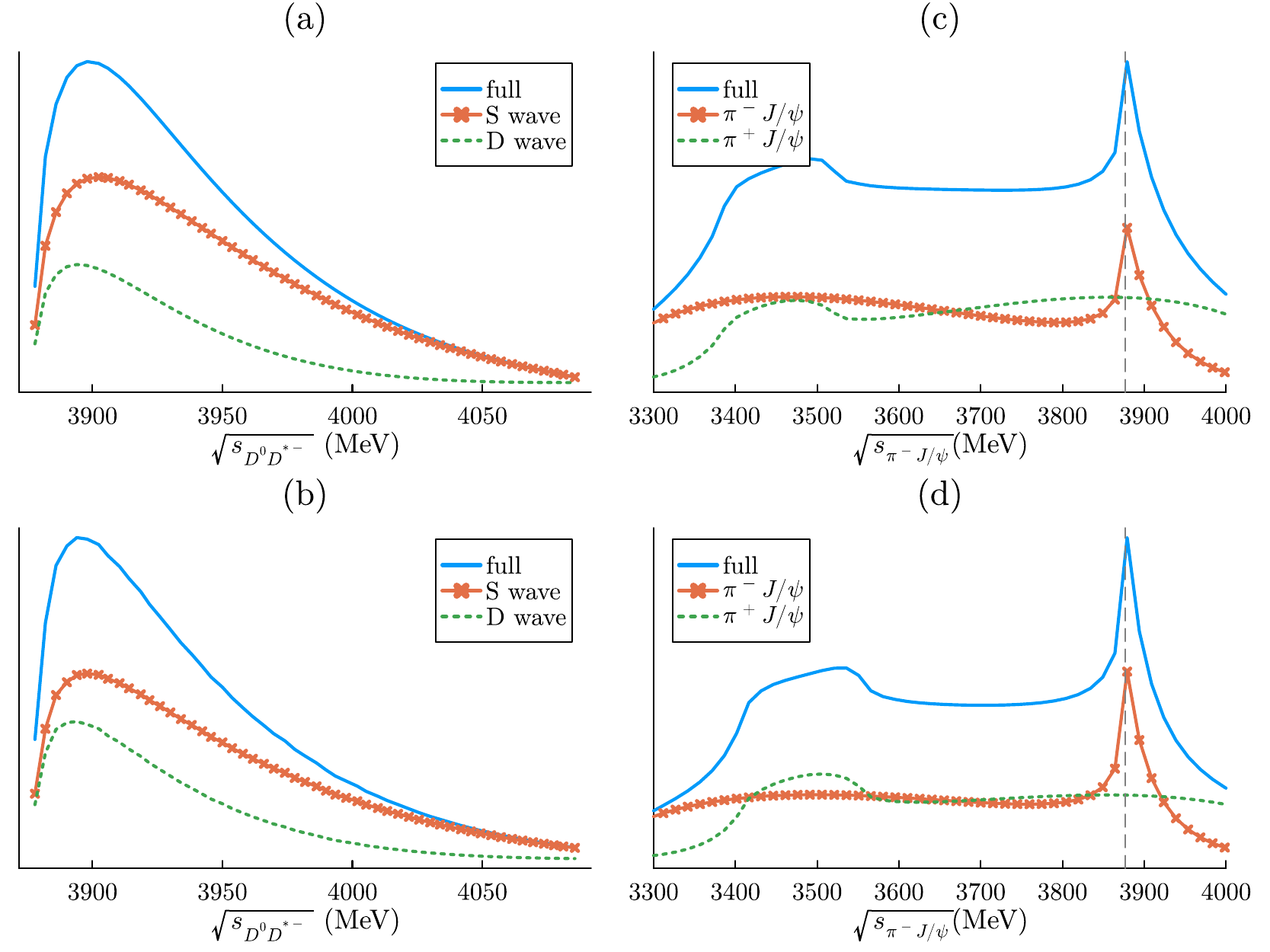}
\caption{Line shapes produced by pure triangle loops for (a,b)$D^0D^{*-}$ and (c,d)$\pi^-J/\psi$ distribution at (a,c) $\sqrt{s}=4.23$ and (b,d) $\sqrt{s}=4.26$ GeV. Blue solid lines, orange lines with markers, green dotted lines correspond to Fig.\ref{fig: feyn diagram pijpsi}(d,e), only Fig.\ref{fig: feyn diagram pijpsi}(d) and only Fig.\ref{fig: feyn diagram pijpsi}(e), respectively, for (a,b) and Fig.\ref{fig: feyn diagram DD}(d), Fig.\ref{fig: feyn diagram DD}(d) with only $S$-wave $D_1\to D^*\pi$ and Fig.\ref{fig: feyn diagram DD}(d) with only $D$-wave $D_1\to D^*\pi$, respectively, for (c,d). The gray dashed vertical lines indicate the threshold of $D\bar{D}^*$. The peak at $\sqrt{s}=4.26$ GeV is slightly narrower since it is closer to the region where TS occurs.
}

\label{fig:puretriangle}
\end{figure}

To investigate the origin of the peak from the triangle diagram, we first present the line shape produced by a pure triangle loop in Fig.\ref{fig:puretriangle}(c,d), as we did earlier. There is a cusp exactly at the threshold, which arises from a branch cut that begins at the threshold $ m_D + m_{D^*}$.
Specifically, when $\sqrt{s}_\alpha \geq m_D + m_{D^*}$ the factor $\sqrt{s_\alpha} - \omega_\pi - \omega_{D^*} - \omega_D$ in the denominator can vanish over some interval of the integration variable. 
In Appendix~\ref{append:triangle}, more details are provided.
Given that the ratio of the height of the cusp to the smooth platform is much smaller than that in Fig.~\ref{fig: fitting lineshape 1}, a $T$-matrix is still necessary, similar to the $D\bar{D}^*$ case. 

\begin{figure*}
\includegraphics[width=0.3\linewidth]{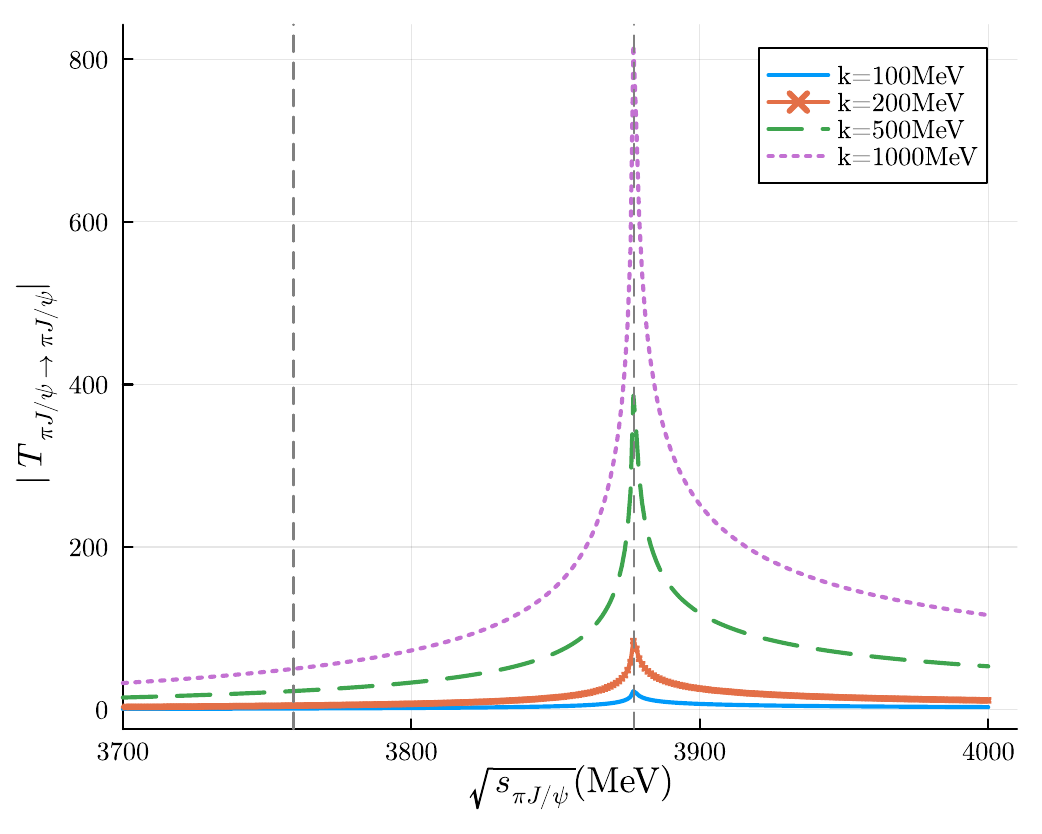}
\includegraphics[width=0.3\linewidth]{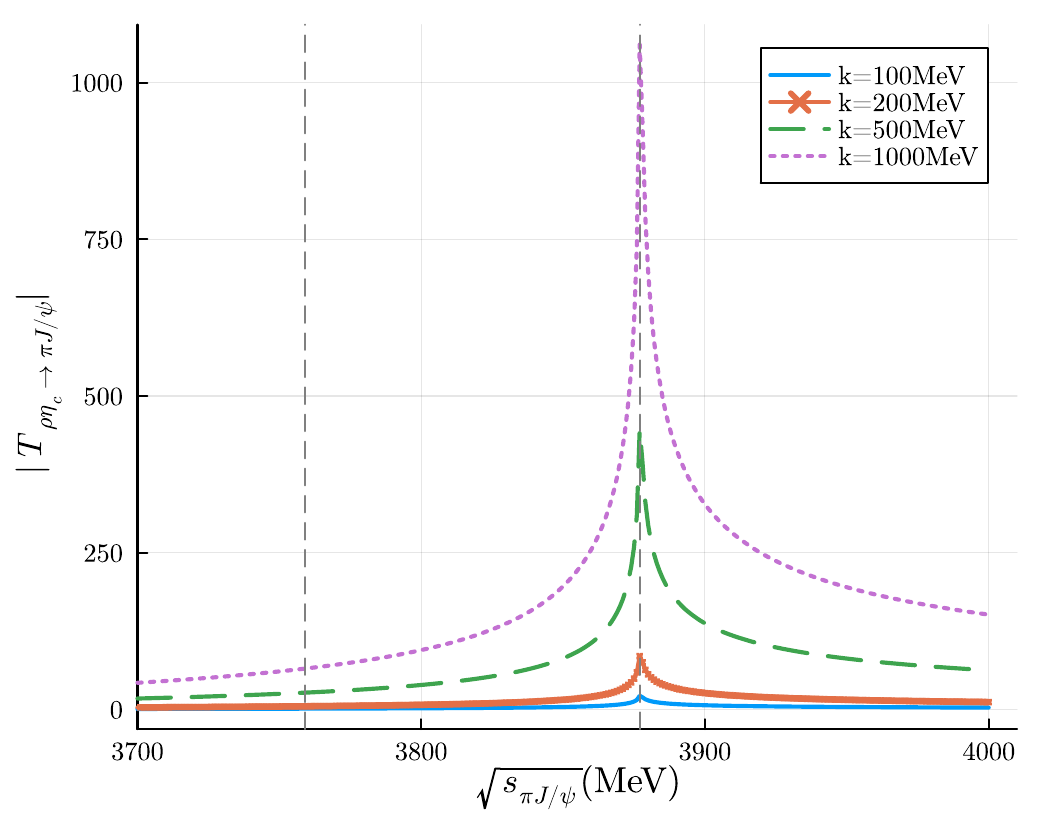}
\includegraphics[width=0.3\linewidth]{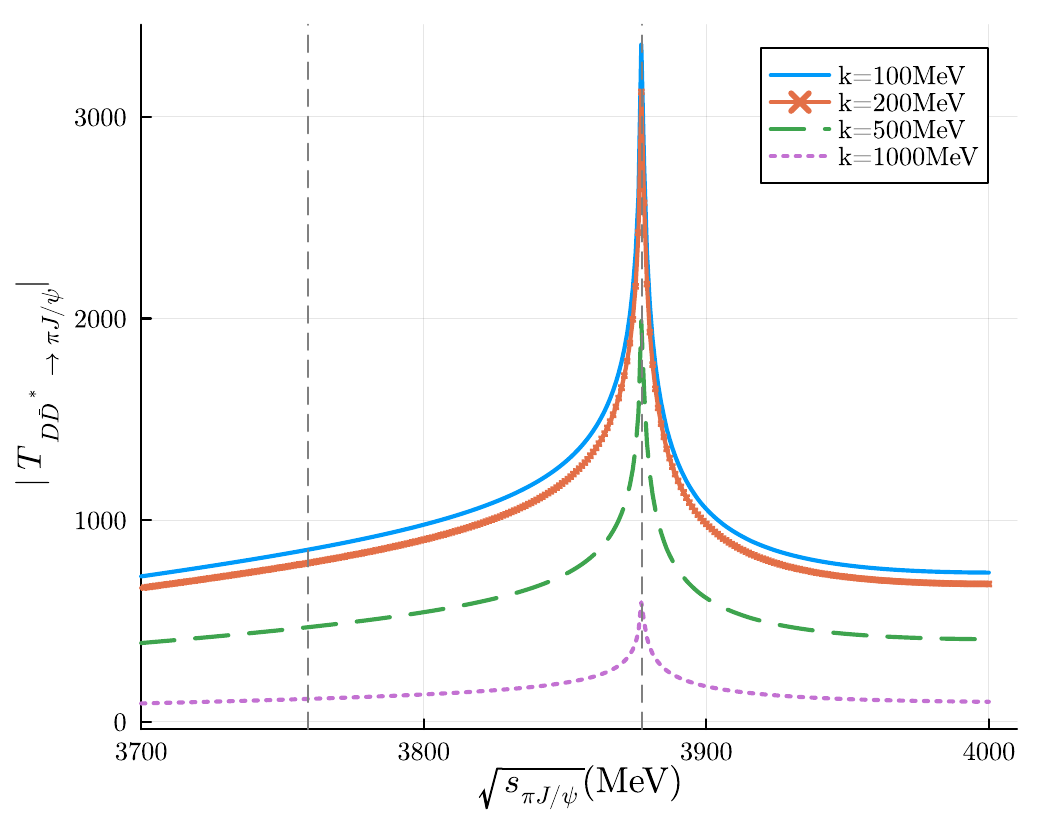}
\caption{
The half-on-shell $\abs{T_{\alpha\to\pi J/\psi}(k,\bar{p}_{\pi J/\psi};\sqrt{s_{\pi J/\psi}})}$ as a function of $\sqrt{s_{\pi^-J/\psi}}$ with fixed off-shell momentum $k$. From left to right $\alpha = \pi^- J/\psi,\rho^-\eta_c$ and $[D\bar{D}^*]^-$. The gray dashed vertical lines denote the threshold of $\rho\eta_c$ and $D\bar{D}^*$. A clear cusp appears at the $D\bar{D}^*$ threshold due to the one-pion-exchange interaction between $D\bar{D}^*$ and $D\bar{D}^*$. 
}
\label{fig: Tmat}
\end{figure*}

Based on the above discussion, the contribution from the $T$-matrix plays a crucial role here so it deserves further investigation. In Fig.~\ref{fig: Tmat}, we illustrate how the magnitude of the half-on-shell $T$-matrix, $\abs{T_{\alpha\to\pi J/\psi}(k,\bar{p}{\pi J/\psi};\sqrt{s{\pi J/\psi}})}$, evolves as a function of $\sqrt{s_{\pi J/\psi}}$, with the loop momentum $k$ fixed at several values.
Due to the one-pion-exchange interaction in the $D\bar{D}^* \to D\bar{D}^*$ process, a noticeable cusp appears at the $D\bar{D}^*$ threshold, stemming from the unitary branch cut. It is important to note that this peak originates from a cusp, not from a resonance or (virtual) bound state, which will be discussed later.
In this context, we assert that the $Z_c$ structure in the $\pi J/\psi$ distribution is a $D\bar{D}^*$ threshold cusp, driven by the $T_{D\bar{D}^*\to \pi J/\psi}$ in the triangle loop involving the $D_1$ particle. 
This conclusion aligns with the findings of the HALQCD collaboration.

Lastly, regarding the $\eta_c \rho$ channel, although its peak signal in the current experimental data is not as prominent as the others, it can also be interpreted as a threshold cusp within our model.

It is interesting that the major contributions to the peaks in the $\pi J/\psi$ and $D\bar{D}^*$ distributions are different in our model. 
This disparity may help explain the small deviation in the central values of the Breit-Wigner parameters for $Z_c(3885)$ and $Z_c(3900)$. 
Furthermore, our model may also provide a possible explanation for why the $Z_c(3900)/Z_c(3885)$ has been observed only in $e^+e^-$ production but not in $B$-decays thus far.

From the above discussion, it is clear that the quality of experimental data, particularly those near the $D\bar{D}^*$ threshold in each channel, is crucial for a deeper and more accurate understanding of the $Z_c$ states. Therefore, high-statistics data from the future STCF experiments~\cite{Achasov:2023gey} will be essential.

\begin{table}
\centering
\begin{tabular}{c|c|c|c}
\toprule
& Scheme &  $\Lambda_{\pi J/\psi(\rho\eta_c)}$ (fixed) & $g^\prime \, (10^{-5}\,\text{MeV}^{-\frac{3}{2})}$    \\ \midrule
\multirow{5}{*}{This work} & \multirow{3}{*}{1}    &  1.3 GeV  &  $7.506 \pm 0.120$   \\
&     &  1.5 GeV  &  $6.143 \pm 0.070 $  \\ 
&     &  1.7 GeV  &  $ 5.129 \pm 0.070$  \\ 
&  \multirow{2}{*}{2}   &  1.5 GeV   &  $5.584 \pm 0.470 $  \\ 
&   &  1.7 GeV  &  $4.69 \pm 0.280$    \\ \midrule
\multirow{2}{*}{Other refs} & -   & - &$3.68$~\cite{He:2017lhy} \\ 
& -  & - & $3.85$~\cite{Yamaguchi:2019djj}  \\ 
\bottomrule
\end{tabular}
\caption{ Fitting results for $g^\prime$ at different fixed cutoff $\Lambda_{\pi J/\psi(\rho\eta_c)}$ for two scheme of OBE potential. The fitting quality for these values are nearly identical, indicating the cutoff-independence of the result. The value of $g^\prime$ in other references are also shown for comparison. The difference can be attributed to the different renormalization scheme. 
}
\label{tab: parameters fit}
\end{table}

\section{Pole Position}\label{sec:pole}

In this study,  three coupled channels result in a total of eight Riemann sheets when the $T$-matrix is analytically continued into the complex plane. The branch cuts of the Green function $G_\alpha$, which extend from the threshold of the $\alpha$-channel to infinity along the positive real  energy axis, connect different Riemann sheets.
For the purpose of analytical continuation, two types of Green functions are defined as follows:
\begin{align}
G^{+}_\alpha(z) &= G_\alpha(z),
\\
G^{-}_\alpha(z) &= G_\alpha(z) - \frac{i}{16\pi^2 z} \bar{p}_\alpha(z) \quad,\Im\bar{p}_\alpha<0.
\end{align}

The Riemann sheets adjacent to the physical region are $\text{RS}^{+++}$, $\text{RS}^{-++}$, $\text{RS}^{--+}$, and $\text{RS}^{---}$, where the superscripts indicate the sign of $G^{\pm}$ for the $\pi J/\psi$, $\rho\eta_c$, and $D\bar{D}^*$ channels, respectively. 
Only one pole is found in $\text{RS}^{---}$, and we examine its dependence on the cutoff and the regularization scheme, as presented in Table \ref{tab: pole position}.

The identified pole is located far below $D\bar{D}^*$ threshold on $\text{RS}^{---}$. For comparison, we list the results from other references that report either virtual or resonant poles in Table ~\ref{tab: pole position}. Since the pole found here is far from the threshold, the $Z_c$ structure near the threshold is primarily due to the threshold cusp effect rather than a resonance or virtual bound state in our model. In Refs. \cite{Du:2022jjv, Albaladejo:2015lob}, a pole located about $80$ MeV below the threshold, which is similar to the one found here, is observed when an energy-independent interaction is adopted. Additionally, the HALQCD Collaboration identifies three poles significantly below the $D\bar{D}^*$ threshold on $\text{RS}^{---}$ and suggests that $Z_c$ is not a conventional resonance but rather a cusp effect, based on their lattice setup \cite{HALQCD:2016ofq}.
Although the value of $m_\pi$ used in LQCD calculations is unphysical and further investigations are necessary, all current lattice data from different collaborations have so far shown no direct evidence of a $Z_c$ resonance~\cite{Prelovsek:2013xba, Prelovsek:2014swa, Chen:2014afa, Cheung:2016bym, HALQCD:2016ofq}.

\begin{table}[]
\centering
\begin{tabular}{c|c|c|c}
\toprule\midrule
& Pole Position  & Type & Scheme($\Lambda_{\pi J/\psi}$) \\ \midrule
\multirow{5}{*}{This work}   & 3798.72 - 1.10i  & \multirow{5}{*}{Virtual} & 1(1.3GeV) \\ 
& 3798.46 - 1.71i & & 1(1.5GeV) \\ 
& 3798.12 - 2.26i  & & 1(1.7GeV) \\ 
& 3798.27 - 2.02i & & 2(1.5GeV) \\ 
& 3797.80 - 2.64i & & 2(1.7GeV) \\ 
\midrule
\multirow{2}{*}{Ref.~\cite{Du:2022jjv}} & $3798\aerror{25}{31}$  & Virtual & \multirow{10}{*}{-} \\
& $3902(6) -  38(9) i$ & Resonance & \\ 
\multirow{2}{*}{Ref.~\cite{Albaladejo:2015lob}} & $3831\aerror{27}{38}$  & Virtual &  \\  
& $3894(6) - 30(13) i$ & Resonance &  \\ 
Ref.~\cite{He:2017lhy} & $3870$ & Virtual &  \\ 
Ref.~\cite{Gong:2016hlt} & $3879$ & Virtual &  \\ 
Ref.~\cite{Ortega:2018cnm} & 3872 & Virtual &  \\ 
Ref.~\cite{Chen:2023def} & 3880(3) - 13(1)i & Resonance & \\ 
Ref.~\cite{vonDetten:2024eie} & 3884 - 22i & Resonance  &  \\
Ref.~\cite{Lin:2024qcq} &  3840 & Virtual &  \\ 
\bottomrule
\end{tabular}
\caption{
The pole found on $\text{RS}^{---}$ for different schemes and cutoffs, far away from the observed peak and therefore has minimal impact. For comparison, the position and type of the pole reported in other references are also shown. 
}
\label{tab: pole position}
\end{table}
%
%
%
%
%
%
\section{Finite Volume Energy Levels}
\label{sec:finitevol}

In this section, we calculate the finite volume energy levels using the HEFT and compare them with the spectra extracted from LQCD. HEFT is a formalism that links finite energy levels to infinite volume scattering amplitudes  and  is equivalent~\cite{Wu:2014vma} to the standard Lüscher formula~\cite{Luscher:1990ux,Hansen:2012tf,Rummukainen:1995vs}.

We begin with a brief introduction. When the system is confined a finite volume with imposed boundary conditions, the momentum becomes discretized. The symmetry group is reduced from the rotational group $O(3)$ to  either the octahedral  group $O_h$ or $C_{2/3/4v}$ depending on the total momentum~\cite{Li:2019qvh,Li:2021mob,Li:2024zld}. Consequently, the potential matrix must be defined in the discretized momentum basis.  Using the $O_h$ symmetry, the Hamiltonian matrix can be block-diagonalized, allowing the extraction of finite volume energy levels corresponding to each irreducible representation (irrep) 
 of $O_h$.  These energy levels can then be compared with the lattice spectrum. This formalism has been successfully applied to describe various systems, such as $\Delta\to N\pi$\cite{Hall:2013qba}, $N^*(1440)$~\cite{Liu:2016uzk, Wu:2017qve}, $N^*(1535)/N^*(1650)$\cite{Liu:2015ktc, Abell:2023nex}, and the positive parity $D_{s}$ states~\cite{Yang:2021tvc} on the lattice.

Given that the spin-parity  $J^P$ of $Z_c$ is reported to be $1^+$ by BES\uppercase\expandafter{\romannumeral3}, it sufficient to calculate the finite volume energy levels for the $T_1^+$ irrep of $O_h$, which is the only representation that couples to $1^+$~\cite{Bernard:2008ax}. The potential matrix for $T_1^+$ is given below. For a general derivation, we refer to Ref.~\cite{inpreparation}.
\begin{align}
\hat{V}^{T_1^+}_{L;\alpha,\beta} = \frac{(2\pi)^3}{L^3} \sum\limits_{N,N^\prime=0}^{N_\text{cut}}\sum\limits_{F=1}^{\text{oc}(T_1^+,N)}\sum\limits_{F^\prime=1}^{\text{oc}(T_1^+,N^\prime)}  \sum\limits_{\boldsymbol{n}}^{\boldsymbol{n}^2=N}\sum\limits_{\boldsymbol{n}^\prime}^{\boldsymbol{n}^{\prime2}=N^\prime}
\notag\\
\frac{V_{\alpha\beta}\left(\boldsymbol{p_{n^\prime}}\, \sigma^\prime,\boldsymbol{p_n}\,\sigma\right) }{(2\pi)^3 4\sqrt{\omega_{\alpha_1}(\boldsymbol{p_{n^\prime}}) \omega_{\alpha_2}(\boldsymbol{p_{n^\prime}})\omega_{\beta_1}(\boldsymbol{p_{n}}) \omega_{\beta_2}(\boldsymbol{p_{n}})  } }
\notag\\
X^{T_1^+}_F(\boldsymbol{n}\,\sigma)X^{T_1^+}_{F^\prime}(\boldsymbol{n}^\prime\,\sigma^\prime) \ket{N^\prime; T_1^+, F^\prime}\bra{N; T_1^+ , F}\,,
\end{align}
where $\boldsymbol{p}_{\boldsymbol{n}} \equiv \frac{2\pi}{L} \boldsymbol{n}$ and $V_{\alpha\beta}$ is the potential in infinite volume. 
The vector $\ket{N;T_1^+,F}$ is orthonormalized and  furnishes $T_1^+$-irrep in the invariant subspace $\mathrm{span}\{\ket{\boldsymbol{p_n}\,,\sigma}|\, \boldsymbol{n}^2=N\}$,  where the occurrence of $T_1^+$ is denoted by $\text{oc}(T_1^+,N)$. 
To construct these vectors, one firstly get $\text{oc}(T_1^+,N)$ linearly independent states using the representation matrix $D^{T_1^+}$ 
\begin{align}
\sum\limits_{R\in O_h} D^{T_1^+}_{1\mu}(R)^*\, R \, \ket{\boldsymbol{p}_{\boldsymbol{n}_0},\sigma} \quad, \boldsymbol{n}_0^2 = N, 
\end{align}
by selecting different $\mu$, $\boldsymbol{n}_0$ and followed by orthonormalization via, for example, the Gram-Schmidt process. 
The function $X^{T_1^+}_F(\boldsymbol{n}\,,\sigma)$ represents the inner product $\braket{N;T_1^+,F}{\boldsymbol{p}_{\boldsymbol{n}_0}\,,\sigma}$ and depends on the orthonormalization. 
Note that the Hamiltonian eigenvalues  are independent of the orthonormalization since there are equivalent up to an unitary transformation. 
While the $\rho$-meson is unstable at physical pion mass and  requires careful  treatment on the lattice, 
it suffices to treat it as  stable here to qualitatively justify our interpretation of $Z_c$.

\begin{figure}
\includegraphics[width=0.9\linewidth]{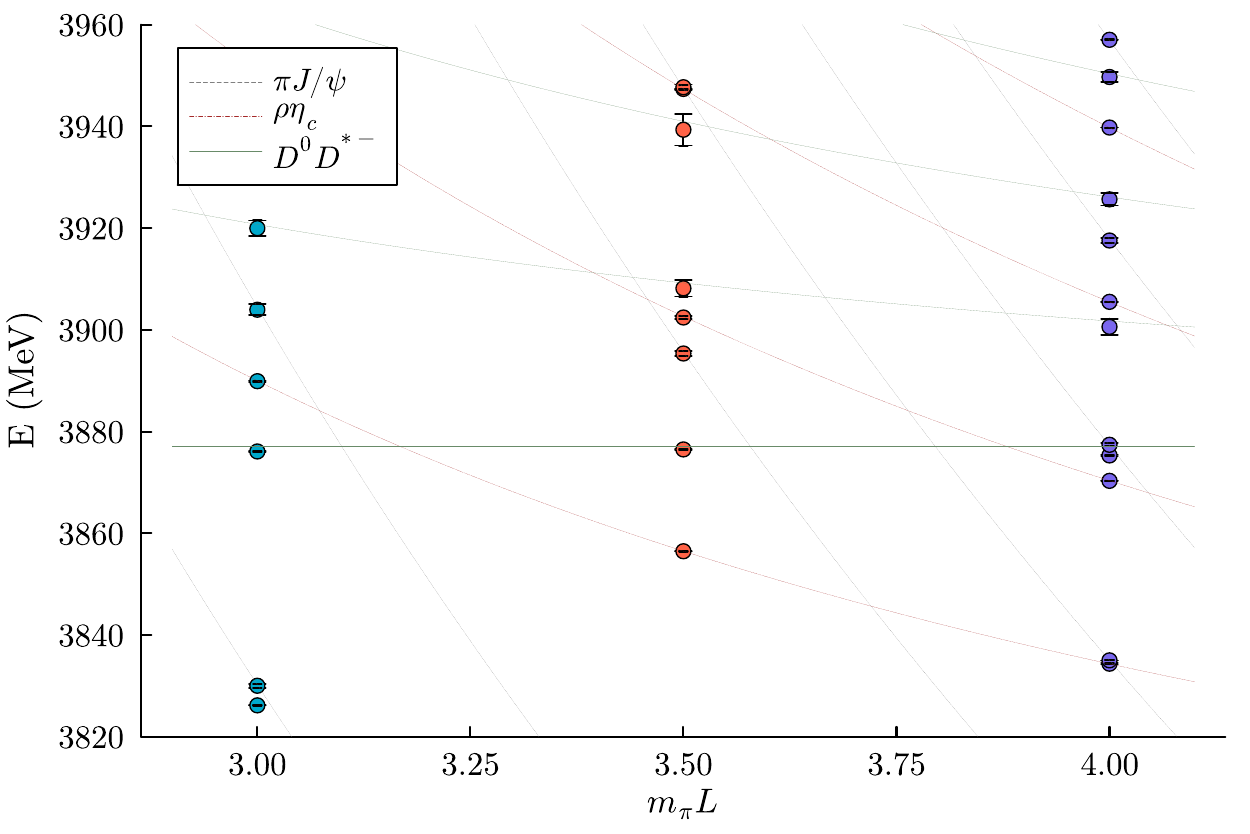}
\caption{Finite volume energy spectra at $m_\pi L = 3,3.5,4$ with $m_\pi=139$ MeV calculated in our model. The circles with error bars represent the energy levels calculated by HEFT. The lines denote the free energy levels of $\pi J/\psi, \rho\eta_c$ and $D\bar{D}^*$.($\rho$-meson are treated as a stable particle here) The energy levels are pretty close to the free energies, indicating the weak interactions.}
\label{fig:FVE}
\end{figure}

The eigenvalues for various box sizes $L$ are shown in the Fig.~\labelcref{fig:FVE}. 
As seen, all  energy levels are close to the free energies of three channels, which indicates that the interactions between them are weak. 
This agrees with the findings given of several LQCD collaboration~\cite{Prelovsek:2013xba, Prelovsek:2014swa, Cheung:2016bym},  despite differences in $m_\pi$. 
Therefore we assert that our model is compatible with the current lattice analyses.

\section{Summary and Prospect}\label{sec:summary}

In this work, a comprehensive coupled-channel model is employed for the first time to analyze the $Z_c$ structure in the $e^+e^- \to \pi^+ (\pi^-J/\psi, D^0D^{*-}, \rho^- \eta_c)$ processes at $\sqrt{s}=4.23$ and $4.26$ GeV, as observed by BES\uppercase\expandafter{\romannumeral3} ~\cite{BES3DD423, BES3DD426, BES3pj423, BES3pj426, Yuan:2018inv}. The OBE interaction that respects heavy quark spin symmetry is adopted to describe the $D\bar{D}^*$ interactions. Additionally, the rescattering $T$-matrix includes a new parameter, $g^\prime$ which governs the coupling vertex between $D\bar D^* $ and charmonium $J/\psi, \eta_c$. Except for $g^\prime$, all other parameters used for the $T$-matrix were determined by previous work, which successfully reproduced the line shape of the double-charmed exotic state $T_{cc}$\cite{Wang:2023ovj}.
Assuming that $e^+e^-$ collisions at $\sqrt{s}=4.23$ and $4.26$ GeV are dominated by $e^+e^- \to Y(4230/4260)$ followed by subsequent decays, the line shapes of the $\pi^- J/\psi$, $D^0D^{*-}$, and $\rho^-\eta_c$ invariant mass distributions are successfully reproduced in our work.

Upon investigating the different contributions, we find that the nature of the $Z_c$ structures in the $D\bar{D}^*$ and $\pi J/\psi$ ($\rho\eta_c$) distributions are not totally the same, which may explain the small deviation between $Z_c(3885)$ and $Z_c(3900)$. In the $D^0 D^{*-}$ distribution, the near-threshold peak primarily arises from the tree-level cascade decay $Y(4230/4260) \to \bar{D}_1(2420)D \to \pi D \bar{D}^*$, which is further enhanced by the triangle diagram that includes the final state interaction (FSI).
In contrast, the $\pi^- J/\psi$($\rho^-\eta_c$) distribution features two cusps. One from the $T$-matrix and the other from the pure triangle loop integral. Both of them occur exactly at the $D\bar{D}^*$ threshold and plays a primary role on the peak structure. Therefore, the $Z_c$ observed in this distribution is identified as a threshold cusp. This interpretation provides a possible explanation to why $Z_c(3900)$ is observed only in $e^+e^-$ collisions and not in $B$-decays, where the production of the $Y$ state is more challenging.

Furthermore, we search for poles of the $T$-matrix on the four Riemann sheets adjacent to the physical region and find only one pole approximately 80 MeV below the $D\bar{D}^*$ threshold on $\text{RS}^{---}$. However, due to its distant location from the threshold, this pole has little impact on the $Z_c$ structure. 

Additionally, we calculate the finite volume energy spectra for the $T_1^+$ irreducible representation, the only one that couples to $J^P=1^+$. All the energy levels in the relevant region we calculate are very close to the free energies due to the weak interactions between channels. This result is qualitatively consistent with the current findings from several different LQCD collaborations~\cite{Prelovsek:2013xba, Prelovsek:2014swa, Cheung:2016bym}, further supporting our interpretation of the $Z_c$ structure. To be more specific, no signal would be observed in the two-body correlation from lattice simulations if $Z_c$ is indeed from a threshold cusp effect.

Based on the results of this work, we conclude that the $Z_c$ structure near 3.9 GeV in the three distributions is due to the threshold cusp arising from coupled channels. However, this conclusion still faces some uncertainties. For instance, we acknowledge that the $\pi\pi$ final state interaction (FSI) is not rigorously treated here, which prevents us from accurately reproducing the experimental $\pi\pi$ invariant mass distribution in the $e^+e^- \to \pi\pi J/\psi$ process. 
In Ref. \cite{Chen:2023def}, the $\pi\pi$ FSI is carefully incorporated using a dispersion relation. Compared to their previous work \cite{Du:2022jjv}, the pole is found on the same Riemann sheet but its position shifts by approximately $20$ MeV. Therefore, to achieve a more reliable and comprehensive analysis, it will be necessary to incorporate the $\pi\pi$ FSI in a self-contained manner in future research. We also suggest that experimental researchers improve the quality of data, especially near the $D\bar{D}^*$ threshold to further elucidate the nature of the $Z_c$.

\section*{Acknowledgements}
The authors want to thank the useful discussions with Changzheng Yuan, Feng-Kun Guo, Chuan Liu and Bing-Song Zou. 
This work is supported by the National Natural Science Foundation of China under Grant Nos. 12175239, 12221005, and 12275046,
and by the National Key Research and Development Program of China under Contracts 2020YFA0406400,
and by the Chinese Academy of Sciences under Grant No. YSBR-101, 
and by the KAKENHI under Grants No.23K03427 and No.24K17055,
and by the Xiaomi Foundation / Xiaomi Young Talents Program

\appendix

\section{Numerical solution of LSE}\label{append:LSE}
The Lippmann-Schwinger equation in the partial-wave representation can be solved numerically. For simplicity, all  indices labeling coupled channels and angular momentum are omitted here. Our goal is to solve the following one-dimensional integral equation
\begin{widetext}
\begin{align}
    T(k^\prime,k;E) = V(k^\prime,k) + \int^{\infty}_0 q^2  dq V(k^\prime,q)G(q;E+i\epsilon)T(q,k),
\end{align}
\end{widetext}
where the infinitesimal $i\epsilon$ can be treated explicitly,
\begin{widetext}
\begin{align*}
    T(k^\prime,k;E) = V(k^\prime,k) + \mathcal{P} \int q^2 dq V(k^\prime,q)G(q;E) T(q,k;E)
    -  i\pi \bar{q} \mu(E) V(k^\prime,\bar{q}) T(\bar{q},k;E)\theta(E-m_1-m_2)
\end{align*}
\end{widetext}
where $\bar{q}$ is defined by $\omega_1(\bar{q}) + \omega_2(\bar{q}) = E$, $\mu(E) = \frac{1}{(2\pi)^3 4E}$ and $\theta$-function vanishes unless $E$ is above the threshold. The symbol $\mathcal{P}$ denotes the Cauchy principal value of the integral. $\bar{q}$ would be a simple pole if $V$ is non-singular. Notice that
\begin{align}
    \mathcal{P} \int \frac{1}{q^2-\bar{q}^2}  dq = 0.
\end{align}
Thus, it is a free lunch to do a subtraction to regularize the integrand,
\begin{widetext}

\begin{align}
 \mathcal{P} \int q^2 dq   V(k^\prime,q)G(q;E)T(q,k;E)   = 
\int q^2 dq  V(k^\prime,q)G(q;E)T(q,k;E) 
- V(k^\prime,\bar{q}) \bar{q}^2 \frac{-2\mu(E)}{q^2-\bar{q}^2} 
 T(\bar{q},k;E)
\end{align}
\end{widetext}
To apply the Gauss–Legendre quadrature method,  the integration domain must first be transformed into a bounded interval. This is done by splitting the domain $(0,\infty)$ into $(0,\hat{q})$ and $(\hat{q},\infty)$, where $\hat{q}$ is an arbitrary positive constant. The integrals over these intervals for any function $f$ can be rewritten  as follows:
\begin{align}
    \int_0^{\hat{q}}  f(q) dq &= \hat{q} \int_0^1  f(x\hat{q}) dx
    \\
    \int_{\hat{q}}^\infty f(q) dq &= \hat{q} \int_0^1 \frac{1-b}{(1-x)^2}f\left(\frac{1-bx}{1-x}\hat{q}\right) dx \,,\forall b<1
\end{align}
Given $N$-points nodes $x_i$ and weights $w_i$ of Gauss–Legendre quadrature on interval $(0,1)$, we define the following quantities
\begin{align}\label{eq:numericalLSEq}
    q_i = \begin{cases}
    \bar{q}(E) & i=0 \\
    x_i \hat{q} &  1\leq i \leq N \\ 
    \frac{1-b x_{i-N}}{1-x_{i-N}}\hat{q} & N<i\leq2N
    \end{cases},
\end{align}

\begin{align}
    \tilde{G}_0(E) &= \frac{q_0^2 \hat{q}}{2E}\left( \sum\limits_{i=1}^N \frac{w_i}{q_i^2-q_0^2} + \sum\limits_{i=N+1}^{2N} \frac{1-b}{(1-x_{i-N})^2} \frac{w_i}{q_i^2-q_0^2} \right) 
    \notag\\
    &\quad- \frac{i\pi q_0}{4E}\theta(E-m_1-m_2)
    \\
    \tilde{G}_i(E) &= \begin{cases}
      w_i q_i^2  \hat{q} \,  G(q_i;E) & 1\leq i \leq N \\
      w_i q_i^2 \frac{1-b}{(1-x_{i-N})^2} \hat{q} \, G(E;q_i) & N<i\leq 2N
      \label{eq:numericalLSEG}
    \end{cases}
\end{align}
and $T_{ij}=T(q_i,q_j), V_{ij}=V(q_i,q_j)$. Consequently, Eq.~\eqref{eq:partialwaveLSE} can be converted into the algebraic equation,
\begin{align}
    T_{ij}(E) = V_{ij} + \sum\limits_{k=0}^{2N} V_{ik}\,\tilde{G}_k(E)\,T_{kj}(E)
\end{align}
or in a matrix notation,
\begin{align}
    [T] = [V] + [V][\tilde{G}][T] .
\end{align}
The solution is given by
\begin{align}
    [T] = (1-[V][\tilde{G}])^{-1} [V]
\end{align}
The on-shell $T$-matrix $T_{00}$ can  be obtained directly. For half-on-shell or off-shell $T$-matrix, given any off-shell momentum $p$, it only needs to extend the definitions in Eq.~\eqref{eq:numericalLSEq} and Eq.~\eqref{eq:numericalLSEG} by introducing $q_{2N+1}=p$ and set $G_{2N+1}=0$.

\section{Numerical evaluation of the triangle loop}\label{append:triangle}
In this section we introduce how to numerically evaluate the Eq.~\eqref{eq:triangle}. For convenience, we introduce a simplified notation,
\begin{align}
    \mathcal{Q} = \int_0^\infty dp \int_{-1}^1 dx \frac{g(p,x;\lambda)}{A(p;\lambda) - \sqrt{B(p;\lambda) + C(p;\lambda)x} + i0^+},
\end{align}
where $\lambda$ denotes the external kinematic variables $\sqrt{s}$ and $\sqrt{s_\alpha}$, which are fixed during the integration. The functions in the denominator are given by,
\begin{align}
    A &=  \sqrt{s} - \sqrt{p_\pi^2 + m_\pi^2} - \sqrt{p^2+m_{D^*}^2} 
    \\
    B &= p^2 + p_\pi^2 + m_D^2 
    \\
    C &= 2pp_\pi
\end{align}
Given $p$ and $\lambda$,  the integrand may has a pole if $A > 0$ and $\abs{\frac{A^2-B}{C}}<1$ with  $x\in\left[-1,1\right]$. Therefore, the integral with respect to $x$ requires careful treatment. A straightforward calculation shows that the integrand is regular when  $s_\alpha +  m_{D^*}^2 - m_D^2 < 2m_{D^*}\sqrt{s_\alpha}$, or equivalently, $\sqrt{s_\alpha}<m_D+m_{D^*}$.  In this case, the integral can be numerically evaluated using a two-dimensional Gauss quadrature method. However, when $\sqrt{s_\alpha}>m_D + m_{D^*}$, the domain of $p$, $(0,\infty)$, should be split into two parts $R$ and $R^\perp$. Here $R$ is the domain where $A>0$ and $\abs{\frac{A^2-B}{C}}<1$, while $R^\perp$ is its complement. Without showing details we find \begin{align}
R=\left[\abs{p_-},\min\left(p_+,\sqrt{\left(\sqrt{s}-\omega_\pi\right)^2 - m_{D^*}^2}\right)\right],
\end{align}
where
\begin{align}
    p_\pm = \frac{\bar{s}_\alpha}{s_\alpha}\left[
    \left(\sqrt{s} - \omega_\pi\right)\sqrt{\frac{1}{4} - \frac{{s}_\alpha}{\bar{s}_\alpha} m_{D^*}^2  } 
    \pm \frac{1}{2}p_\pi
    \right] .
\end{align}
In the range of $\sqrt{s_\alpha}$ considered, $\min(\cdots)\equiv p_+$. When $p \in R^\perp$, the two-dimensional Gauss quadrature formula can be applied directly.  For $p \in R$, the $i0^+$ prescription should be worked out manually as
\begin{align}
    \mathcal{Q} &\supset \int_R dp \left(\mathcal{P}\int_{-1}^1  \frac{g(p,x)dx}{A-\sqrt{B+Cx}} - 
    i\pi \frac{2A}{C} g(p,x_0) \right) \notag
    \\
    &= \int_R dp \left[ \mathcal{P}\int_{-1}^1\left(\frac{g(p,x)dx}{A-\sqrt{B+Cx}} + \frac{2A}{C}\frac{g(p,x_0)}{x-x_0}\right)dx 
    \right. \notag
    \\
    & \left. -\frac{2A}{C} g(p,x_0) \log\left(\frac{1-x_0}{1+x_0}\right) -i\pi \frac{2A}{C} g(p,x_0) \right] \notag
    \\
    &= \int_R dp \int_{-1}^1\left[  \frac{g(p,x)}{A-\sqrt{B+Cx}} + \frac{2A}{C}g(p,x_0) \left(\frac{1}{x-x_0} 
    \right.\right.  \notag
    \\
    &
    \qquad\qquad\left.\left. - \frac{1}{2}\log\left(\frac{1-x_0}{1+x_0}\right) - \frac{1}{2}i\pi\right)\right]dx,
\end{align}
where $x_0=\frac{A^2-B}{C}$. In the last equation, the symbol $\mathcal{P}$ is dropped since the integrand becomes regular after subtracting a function with the same pole structure as the original integrand. This subtraction enables  numerical evaluation using the two-dimensional Gauss quadrature formula. Note that we here adopt $\frac{1}{x-x_0}$ for subtraction instead of $\frac{1}{x^2-x_0^2}$ which would introduce an superfluous pole at $x=-x_0$ within interval $(-1,1)$.

\section{Potential}\label{append:potential}
In this section, we present the tree-level amplitude $i\mathcal{M}$ derived from the lagrangians in Eqs.~\eqref{eq:HQETcharomium1} to~\eqref{eq:HQETcharomium5}. Firstly, the propagators of heavy meson and charmonium in terms of residual momentum within the framework of heavy quark effective field theory(HQEFT) are given by:
\begin{align}
    \frac{i}{2 v \cdot k - \frac{\boldsymbol{k}^2}{m_H} + i\epsilon} \approx \frac{i}{(k+m_H v)^2-m_H^2+i\epsilon}m_H^2
\end{align}
for (pseudo-)scalar particles and
\begin{align}\label{eq:HQETvecpropagator}
    \frac{i(-g^{\mu\nu}+v^\mu v^\nu)}{2v\cdot k - \frac{\boldsymbol{k}^2}{m_H} + i\epsilon} \approx \frac{i(-g^{\mu\nu}+v^\mu v^\nu)}{(k+m_H v)^2-m_H^2+i\epsilon}m_H^2
\end{align}
for (pseudo-)vector particles. Secondly, the polarization vectors of heavy mesons and charmonium are treated to be static at leading order, i.e,
\begin{align}
    \epsilon^\mu_0 &= \epsilon^\mu_0(\boldsymbol{0}) =  \left(0,0,0,1\right), \\
    \epsilon^\mu_{\pm1} &=\epsilon^\mu_{\pm1}(\boldsymbol{0}) = \frac{1}{\sqrt{2}}\left(0,\mp1,-i,0\right),
\end{align}
which satisfy the condition $v\cdot\epsilon=0$ and the spin sum $\sum\limits_{\sigma}\epsilon_\sigma^{*\mu}\epsilon^\nu_\sigma$ gives the numerator of Eq.(\ref{eq:HQETvecpropagator})

With these ingredients, the amplitude in terms of full momentum can be obtained. In Ref.~\cite{Wang:2023ovj}, the amplitude $i\mathcal{M}_{[D\bar{D}^*] \to [D\bar{D}^*]}$ has already been provided so we only need to present $i\mathcal{M}_{\alpha(\boldsymbol{k})\to [D\bar{D}^*](\boldsymbol{p})}\,,\alpha=\pi J/\psi,\rho\eta_c$. 
Due to the G-parity symmetry:
\begin{align}
    i\mathcal{M}_{\alpha \to [D\bar{D}^*]} = i \sqrt{2}  \mathcal{M}_{\alpha\to D^0 D^{*-} }\equiv i\mathcal{M}^\alpha
\end{align}
For $\alpha = \pi J/\psi$,
\begin{align}
    \frac{i\mathcal{M}_1^{\pi\psi}}{\sqrt{m_\psi m_{D^*} m_D^3}} &= \frac{i4gg^\prime}{f_\pi}\frac{\left(\boldsymbol{\epsilon}_\sigma\cdot(2\boldsymbol{p}+\boldsymbol{k})\right)\left(\boldsymbol{\epsilon}^*_{\sigma^\prime}\cdot\boldsymbol{k}\right)}{q^2-m_D^2+i\epsilon} 
    \\
    \frac{i\mathcal{M}_2^{\pi\psi}}{\sqrt{m_\psi m_{D^*}^3 m_D}} &= \frac{igg^\prime}{f_\pi}\frac{\left((\boldsymbol{k}\times\boldsymbol{\epsilon}^*_{\sigma^\prime})\cdot(2\boldsymbol{p}+\boldsymbol{k})\times\boldsymbol{\epsilon}_\sigma\right)}{q^2-m_{D^*}^2 + i\epsilon}
    \\
    \frac{i\mathcal{M}_3^{\pi\psi}}{\sqrt{m_\psi m_{D^*}^3 m_D}} &= \frac{i4gg^\prime}{f_\pi}\frac{
    k_i(k-2p)_j
    }{q^2-m_{D^*}^2+i\epsilon} \left(  \delta^{ij}\delta_{\sigma\sigma^\prime}   \right. \notag 
    \\ &  \left. + \epsilon^i_\sigma \epsilon^{j*}_{\sigma^\prime} - \epsilon^j_\sigma \epsilon^{i*}_{\sigma^\prime}\right)
\end{align}
where $g=0.6$ is the decay constant of $D^*$. For $\alpha=\rho\eta_c$, 
\begin{align}
   & \frac{i\mathcal{M}_1^{\rho\eta_c}}{\sqrt{m_{\eta_c}m_{D^*}^3m_D}} = \frac{g^\prime g^\prime_V}{q^2-m_{D^*}^2 + i\epsilon}\left[\beta\epsilon_\sigma^0(\boldsymbol{k}) (\boldsymbol{k}-2\boldsymbol{p})\cdot\boldsymbol{\epsilon}_{\sigma^\prime}^* \right.
   \nonumber \\
  &  + \left. 2\lambda \left(\boldsymbol{k}\cdot(\boldsymbol{k}-2\boldsymbol{p})\delta_{\sigma\sigma^\prime} - (\boldsymbol{k}-2\boldsymbol{p})\cdot\boldsymbol{\epsilon}_\sigma(\boldsymbol{k})\boldsymbol{k}\cdot\boldsymbol{\epsilon}_{\sigma^\prime}^*\right)
    \right]
    \\
&   \frac{i\mathcal{M}_{2}^{\rho\eta_c}}{\sqrt{m_{\eta_c}m_{D^*}m_D^3} } =  \beta g^\prime g^\prime_V  \epsilon^0_\sigma(\boldsymbol{k}) \frac{\left(\boldsymbol{k}+2\boldsymbol{p}\right)\cdot\boldsymbol{\epsilon}_{\sigma^\prime}^*}{q^2- m_D^2 + i\epsilon}
    \\
 &   \frac{i\mathcal{M}^{\rho\eta_c}_3}{\sqrt{m_{\eta_c}m_{D^*}^3m_D}} = 2\lambda g^\prime g^\prime_V \frac{\left(\boldsymbol{k}\times\boldsymbol{\epsilon}_\sigma(\boldsymbol{k})\right)\cdot\left((\boldsymbol{k}+2\boldsymbol{p})\times\boldsymbol{\epsilon}_{\sigma^\prime}^*\right)}{q^2-m_{D^*}^2+i\epsilon}
\end{align}
where $g^\prime_V = 2\sqrt{2} g_V=2\sqrt{2}\times5.8 = 16.4$, $\lambda=0.683$/GeV and $\beta=0.687$ (at $\Lambda_{[D\bar{D}]^*}=1$ GeV) are taken from Ref.~\cite{Wang:2023ovj}. 
$\epsilon^\mu_\sigma(\boldsymbol{k})$ is the polarization vector of the $\rho$ light-meson,
\begin{align}
    \epsilon^0_\sigma(\boldsymbol{k}) &= \frac{1}{m_\rho}\boldsymbol{k}\cdot\boldsymbol{\epsilon}_\sigma(\boldsymbol{0}), \\
    \epsilon^i_{\sigma}(\boldsymbol{k}) &= \boldsymbol{\epsilon}^i_\sigma(\boldsymbol{0}) + \left(\frac{\sqrt{\boldsymbol{k}^2+m_\rho^2}}{m_\rho} -1 \right) \frac{\boldsymbol{k}^i \left(\boldsymbol{k}\cdot\boldsymbol{\epsilon}_\sigma(\boldsymbol{0})\right)}{\abs{\boldsymbol{k}}^2}.
\end{align}

\section{Fitting without $Y\to D_1\bar{D}$ + c.c vertex}
\label{append: no tri}

\begin{figure}[h]
    \includegraphics[width=0.95\linewidth]{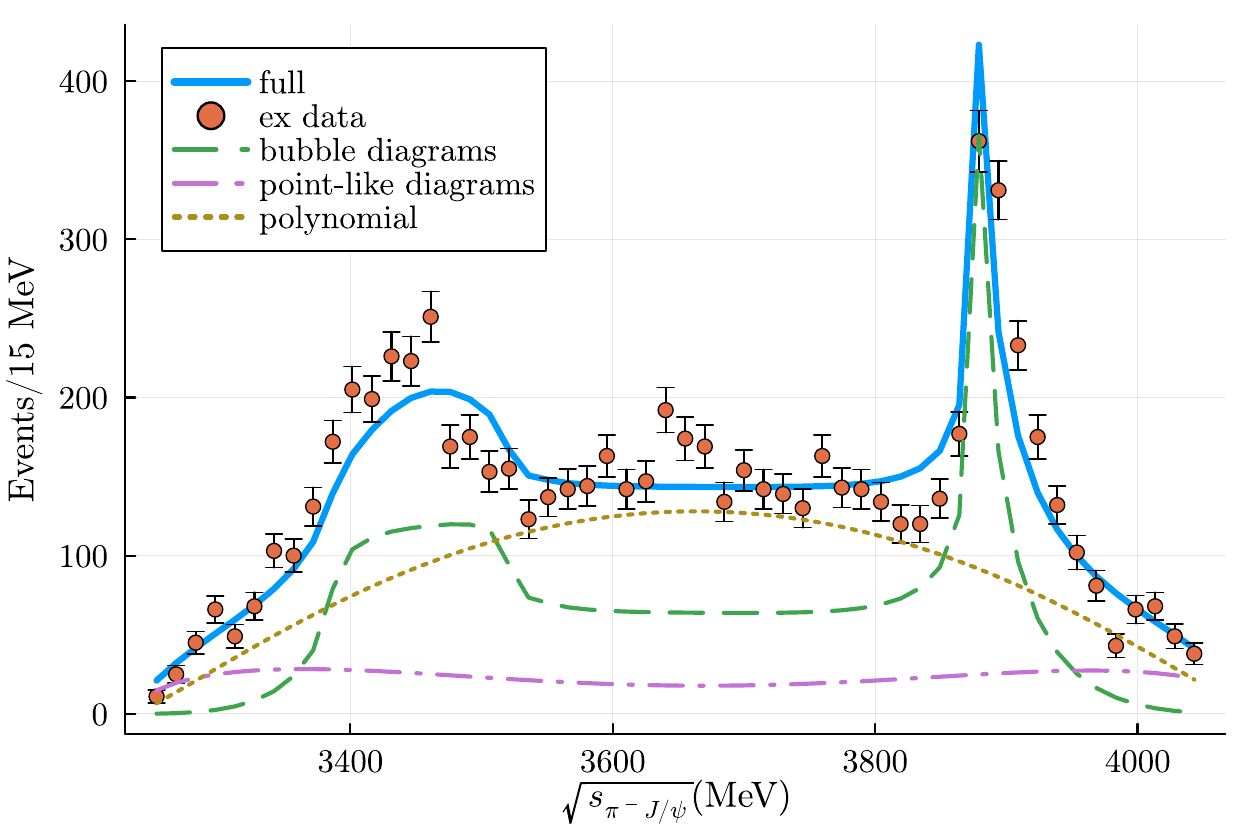}\\
    \includegraphics[width=0.95\linewidth]{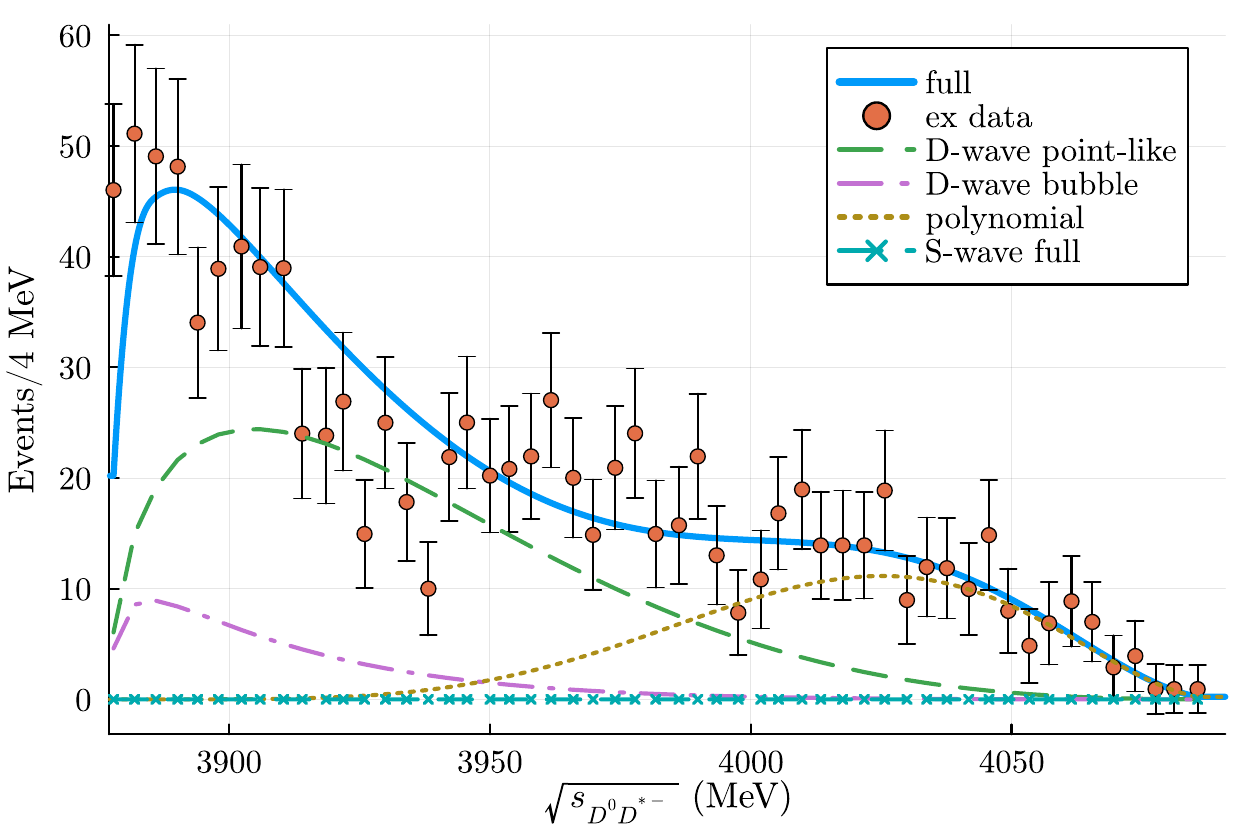}
    \caption{Line shapes determined by fitting the experimental data at $\sqrt{s}=4.23$ GeV without $Y\to D_1\bar{D}$ vertex. The cutoff $\Lambda_{\pi J/\psi(\rho\eta_c)}$ is fixed at $2.0$ GeV. Orange solid circles are experimental data. Blue solid, green dashed pink dot dashed and yellow dotted lines in the top(bottom) row denote the contribution from full amplitude, bubble diagrams($D$-wave $Y\to \pi_D(D \bar{D}^*)$ point-like diagram), point-like diagrams($D$-wave $Y\to \pi_D(D \bar{D}^*)$ bubble diagram) and the polynomial function, respectively. Line with markers in the bottom row denote the highly-suppressed contribution from $S$-wave $Y\to \pi_S(D\bar{D}^*)$ point-like as well as bubble diagram. 
    }
    \label{fig:notri}
\end{figure}

In this appendix we exclude the $Y\to D_1 \bar{D}$ vertex and attempt to fit the experimental data using only point-like and bubble diagrams. Specifically, the cascade decay $Y\to D_1 \bar{D}\to \pi \alpha$ and the triangle diagram are omitted, while the FSI are still considered. 

As discussed in the main text, the peak in $D\bar{D}^*$ distribution  cannot be reproduced using only $S$-wave $Y\to \pi_S(D\bar{D}^*)$ vertex. To address this, we introduce an extra $D$-wave $Y\to \pi_D(D\bar{D}^*)$ vertex, which ultimately lead to a term proportional to $p_\pi^4$ in $\abs{\mathcal{M}}^2$. We present the fitting for $\pi J/\psi$ and $D\bar{D}^*$ distribution at $\sqrt{s}=4.23$ GeV in Fig.~\ref{fig:notri}. 

We now offer some observations. First, the peak in $\pi^- J/\psi$ distribution can indeed be produced if $Y\to D\bar{D}^*\pi$ is significantly much stronger than $Y\to\pi\pi J/\psi$. Second, the peak in $D\bar{D}^*$ distribution can also be obtained with the help of $D$-wave $Y\to \pi_D(D\bar{D}^*)$ vertex. This is easy to understand because $p_\pi$ increases as $\sqrt{s_{D^0 D^{*-}}}$ approaches threshold. However, the contribution from $S$-wave $Y\to \pi_S(D\bar{D}^*)$ is highly suppressed. This is quite unnatural since the phase space of $Y \to D\bar{D}^*\pi$ is not large and the typical $p_\pi$ is small, the high partial wave is expected to be suppressed for naturalness. If the $D$-wave contribution is significantly larger than $S$-wave contribution, it would be difficult to justify ignoring $G$- or even higher partial wave. Lastly, the reduced $\chi^2$ is considerably worse than the fit in the main text. 

In conclusion, though the experiment data can more or less be fitted without $Y\to D_1\bar{D}$ vertex, the result is physically and theoretically unsound. Therefore, this vertex should be included and the triangle diagram must be incorporated into the formalism for self-consistency.

\bibliography{ref}

\end{document}